\documentclass[manuscript,screen,nonacm]{acmart}

\setcopyright{none}
\renewcommand\footnotetextcopyrightpermission[1]{}
\newcommand{\rev}[1]{#1}
\newcommand{\rr}[1]{#1}
\acmISBN{}
\acmDOI{}

\usepackage{booktabs}

\begin{document}

\title{One Tool, One Taste? How Vibe Coding Trades Collective Diversity for Individual Creativity}

\author{L\'eonard Boussioux*}
\email{leobix@uw.edu}
\affiliation{
  \institution{Michael G. Foster School of Business, University of Washington}
  \city{Seattle}
  \state{WA}
  \country{USA}
}
\author{Ziyi Zhao*}
\email{ziyzhao@fiu.edu}
\affiliation{
  \institution{College of Business, Florida International University}
  \city{Miami}
  \state{FL}
  \country{USA}
}
\author{Kanghyun Cho*}
\email{simcho@iu.edu}
\affiliation{
  \institution{Kelley School of Business, Indiana University}
  \city{Bloomington}
  \state{IN}
  \country{USA}
}

\begin{abstract}
Vibe-coding systems turn natural-language instructions into deployed websites, but little is known about the diversity of designs produced within a shared production context. We study 73 promotional websites created by graduate students with Lovable for distinct real businesses under a graded assignment that rewarded original design. We represent each homepage using DINOv3 embeddings and analyze pairwise cosine similarity, effective diversity, and descriptive cluster structure. A representative UMAP--HDBSCAN specification assigns 88\% of sites to six interpretable clusters, while sensitivity analyses yield between three and twelve clusters. In a process subsample of nine recordings, preliminary AI-assisted coding documents recurring generate-and-check behavior, incomplete transmission of verbalized intentions into prompts, and infrequent explicit evaluation of originality. Among survey respondents, perceived human control, satisfaction, and perceived quality show no detectable association with embedding-based cohort atypicality. These results document within-cohort visual concentration and a mismatch between subjective experience and distributional distinctiveness. We call this condition authored ignorance. Alongside early evidence that the collective cost of generative production reaches visual design, we contribute a measurement strategy that joins outcome, process, and perception for interactive artifacts, and design implications running from originality feedback to deliberate friction.
\end{abstract}



\maketitle

\section{Introduction}

Vibe-coding platforms such as Lovable, v0, Bolt, and Replit convert a plain-language description (i.e., ``a warm, welcoming site for a family-run French restaurant'') into a fully functional website within minutes without needing to know how to program. The user can refine the result through further conversation, so that ideation, prototyping, and implementation occur within a single conversational loop \citep{li2026}. Building a professional website once required an agency budget or weeks of skilled effort; now it requires a plain conversation. Adoption has grown quickly: by early 2026, Lovable alone reported roughly eight million users, around 200,000 projects built or updated per day, and more than \$400 million in annual recurring revenue \citep{heim2026}.

A website is one of the main ways a business sets itself apart. Appearance drives recognition and shapes the impression a
business makes on someone who has never encountered it before
\citep{henderson1998, lindgaard2006}. Text-based domains already
show what happens to that differentiation when many producers draw
on one source. Writers supplied with ideas from a large language
model (LLM) produce better stories individually but more similar
stories collectively \citep{doshi2024}. In open-innovation contests, crowds generate more novel solutions than human-AI pairs, even as AI-assisted solutions match or exceed them in value and feasibility \citep{boussioux2024crowdless}. The same pattern recurs across domains: generative AI increases individual performance while narrowing collective diversity \citep{boussioux2026hidden, padmakumar2024, anderson2024}.

Vibe coding intensifies this concern by altering the regime of production. The studies examined AI as an aid to human composition, where the system offered suggestions that the author could accept, reject, or revise, and homogenization operated through anchoring \citep{boussioux2026hidden}. In vibe coding, the generative system produces the entire artifact end-to-end, including layout, palette, typography, imagery, and copy. Functionally, the tool moves from assistant to designer, and the interaction typically reduces to a single loop between a chat box and a live preview. The work of building reorganizes around generation latency and brief, repeated acts of inspection. Each request yields a single finished design, and the platform provides no alternatives for comparison. The most consequential change is who gets to specify: whatever the builder leaves unstated, the production stack settles. Those defaults are the model's learned visual priors together with the platform's templates, components, and themes, and they are the baseline each artifact departs from, if it departs at all.
Because a small number of platforms mediate this production at scale, those defaults are positioned to become a house style for the web. A shift in who decides cannot be read off the outputs alone. It has to be watched inside the loop where the deciding happens, and checked against what the people in that loop believe.

We therefore pose three questions. \emph{RQ1 (outcome):} Does production through a shared generative layer homogenize visual artifacts, even when inputs are divergent, and originality is explicitly incentivized? \emph{RQ2 (mechanism):} If so, through what process does the homogenization operate: where, between a builder's intention and a shipped website, do the production stack's defaults take over? \emph{RQ3 (perception):} Why does the convergence go unnoticed by the very people producing it?

To answer these questions, we leverage a course deployment that observes the same builders at three levels: (i) the artifacts they produced, (ii) the process that produced them, and (iii) their perceptions of both. Seventy-three graduate students each selected a \emph{different} real business (e.g., coffee shops, bakeries, restaurants, nail salons, gyms, professional photographers, NGOs, IT consultancies) and built a promotional website for it on the same vibe-coding platform (Lovable), as a graded course assignment with real stakeholders, since students had to gather feedback from the business and from prospective customers. No two students built for the same business and the grading rubric explicitly rewarded ``creative and original design that avoids generic AI aesthetics.'' The \emph{artifact layer} comprises all published promotional sites, embedded with a self-supervised vision transformer \citep[DINOv3;][]{simeoni2025}, yielding 2,628 pairwise visual comparisons, with a site's \emph{originality} defined as its mean visual distance from the rest of the cohort. The \emph{perception layer} comprises post-task surveys of felt authorship, satisfaction, and perceived quality (58 of 73 builders). The \emph{process layer} comprises nine screen-recorded build sessions with concurrent think-aloud, instrumented to one-second resolution (Section~\ref{sec:process}). 

Four findings, developed across the three results sections, organize the paper.

First, the design space collapses. Unsupervised clustering recovers six recurring visual archetypes, or house styles, that account for 64 of 73 sites (88\%), and the effective number of distinct designs is roughly a dozen. The median builder's nearest neighbor in the cohort shares 49\% visual similarity, against a 23\% average across all pairs. The collapse retains real residual variety, but its scale is notable: 73 independent design problems reduce to roughly a dozen recurring designs.

Second, the collapse co-occurs with a uniform process that supplies its candidate mechanism. Every chat-driven builder runs a similar generate-and-check process, and the same second-by-second reflexes recur around every prompt. Within that script, human input is thin and thinning: a share of voiced intentions never reaches the typed prompt, and what builders do type narrows over the course of a session, from substantive requests to small, generic adjustments. Whatever goes unspecified is decided by the production stack's defaults. The process data measures how large that unspecified space is, down to the keystroke.

Third, the defaults survive because the acceptance test never examines them. Spoken verdicts are fast, generic (\emph{good}, \emph{cool}, \emph{cute}), and frequently absent; originality is invoked in 3.4\% of 440 verdicts. Even visibly flagged defects often ship. Defects and defaults exit the loop by the same route, acceptance without verification.

Fourth, and most consequentially, the convergence is invisible from the inside, and the process data shows the invisibility being produced in real time. Builders narrate the session in the first person while crediting the \emph{making} to the AI, an experience of activity without authorship in the evaluative sense. The survey layer confirms the decoupling at the cohort scale. Students were explicitly instructed and graded to avoid ``generic AI aesthetics''; they nonetheless converged while feeling they had not. Authorship, satisfaction, and perceived quality were all uncorrelated with measured originality. Builders cannot perceive their own position in a distribution they cannot see; only the platform sees the distribution. We name this \emph{authored ignorance}, meaning felt authorship of an artifact whose genericity its author has no way to perceive, because perceiving it would require the distribution of everyone else's artifacts, which only the platform sees.

In sum, whatever the builder does not specify is decided by the production stack's defaults; the shared loop keeps the unspecified space large, and the acceptance test leaves it unexamined. The rest of the paper develops each clause of this account.

This paper makes three contributions. To research on generative AI and creativity, we provide early field evidence on the collective side of the individual-versus-collective tension in visual design produced end-to-end by generative tools, a domain where artifacts are high-dimensional, perceived holistically, and built for real stakeholders. This extends the text-domain diversity findings above, and links them to the pre-generative homogenization of the web documented by \citet{goree2021}. We pair this with a process-level account of how that collective cost is produced, second by second, inside the build loop. To methodology, we contribute a joint outcome--process--perception measurement strategy for interactive artifacts: full-page vision-transformer embeddings, similarity distributions, archetype discovery, and effective-diversity indices on the outcome side, paired with one-second behavioral instrumentation on the mechanism side. To design, we anchor each implication to a specific, documented mechanism rather than a general principle. Originality feedback answers the invisibility of the distribution; divergence-promoting generation targets the single shared script; and deliberate friction is not a new idea, but our process data locates exactly where it would bite and shows it can be funded by time the loop already spends idle.

\section{Background}

The study draws on three literatures, covering evidence on generative AI and collective creativity, the interaction regime that vibe coding establishes, and the computational measurement of visual similarity.

\subsection{Individual Creativity, Collective Diversity}

Creativity research conventionally requires that ideas be both novel and useful \citep{amabile1983}. Novelty is a property of an artifact relative to precedent: whether it departs from what came before, either for its own producer (Boden's psychological creativity) or for the population at large (historical creativity) \citep{boden2004creative}. Atypicality is a property of an artifact relative to a specific reference set: how far it sits from the other members of a defined population, without any claim about precedent or history \citep{uzzi2013atypical}.  Atypicality is not automatically an asset. Typical designs are
liked more on first contact and can sell better for it
\citep{landwehr2011}, and the advantage flips to atypical designs only
under repeated exposure \citep{landwehr2013}. Our claim is therefore
about the cohort's distribution, not about any single site being worse
for sitting near its center; Section~\ref{sec:fluency} returns to this. 

Diversity is a property of the population itself: how spread out, or how many effectively distinct types, its members collectively occupy \citep{jost2006}. A single atypical artifact does not imply a diverse population, since a population can contain one outlier and ninety-nine near-duplicates; nor does a diverse population imply that any one artifact within it is atypical relative to the others.

Generative AI has a documented and growing record on measures of human-AI creativity. In controlled story-writing experiments, access to model-generated ideas raised individual novelty and usefulness, most for the least creative writers, while increasing the similarity among stories written with AI assistance; writers anchored on what the model offered \citep{doshi2024}. In a field crowdsourcing challenge, human solvers produced more novel solutions than human-AI teams, while AI-assisted solutions scored higher on feasibility and value \citep{boussioux2024crowdless}. Parallel results have accumulated for collaborative writing, where the assistance of an instruction-tuned model reduces content diversity \citep{padmakumar2024}, and for ideation, where participants using ChatGPT produced less semantically distinct ideas than users of an alternative creativity support tool \citep{anderson2024}. A synthesis of this work frames the pattern as a hidden cost: AI lifts the individual and narrows the collective, and the narrowing depends on \emph{where} in the creative process the AI enters. Assistance during ideation homogenizes; human-led ideation with AI refinement does not largely \citep{boussioux2026hidden}. At the ecosystem level, the concern parallels \emph{algorithmic monoculture}: when many independent actors adopt the same algorithm because it is individually superior, aggregate outcomes can worsen even as each actor improves \citep{kleinberg2021}. \citet{huang2025} describe the generative version of this trajectory as an \emph{average trap}, where next-token prediction pulls outputs toward generic forms that lack individuality, and where the trap deepens as models learn from their own outputs.

Most of this evidence refers to text. A smaller body extends the pattern to static visual art: adoption of text-to-image generators increased individual artists' novelty and market value while average content novelty across the platform declined as practice standardized \citep{zhoulee2024generative}, and a later study on the same art platform finds that AI-assisted creators contribute more novel work in aggregate through higher output, even as the average novelty of any single piece falls, a productivity effect with no detectable human-AI synergy on top of it \citep{zhou2025hive}. Websites depart from a single generated image in ways that further raise the collective question. A visual artifact is perceived holistically and at a glance. A viewer more quickly and reliably registers two websites as similar than a reader registers two essays as making similar arguments. A business's visual identity also carries economic weight that prose generally does not. For customers deciding in seconds whether to trust an unfamiliar business, the appearance of the website often functions as the pitch itself. Moreover, in vibe coding, the production regime departs from AI-assisted writing. We will discuss this distinction next.

\subsection{Vibe Coding as Delegation}

``Vibe coding,'' a term coined by \citet{karpathy2025} in a post describing a style of development where one ``fully give[s] in to the vibes'' and forgets the code exists, refers to development in which a user expresses intent in natural language and a generative system produces working software. Interview evidence from product teams describes four iterative stages: context setup and ideation, AI generation and refinement, manual debugging and editing, and testing and review \citep{li2026}. The same study records concerns about ``shallow creativity'' and the erosion of distinctive design voices, while related conceptual work maps possible sources of design homogenization and proposes productive friction as a mitigation \citep{shin2026interrogating}. Our study complements this literature by measuring artifact similarity and observing a small set of production sessions. It does not assume that all vibe-coding systems eliminate manual composition: builders can request, inspect, and sometimes directly edit fonts, palettes, layouts, content, and code. The empirical question is how those opportunities were used in this cohort.

Throughout, we use \emph{production-stack defaults} as a broad label for design choices supplied by an underlying model together with any platform scaffolding, such as templates, component libraries, and themes, when a builder leaves those choices open. The present design does not separate model priors, platform scaffolding, general web conventions, instructional inputs, or builder preferences; doing so requires controlled comparisons across tools and models (Section~\ref{sec:limits}).

This framing suggests two possibilities. First, open-ended prompts may leave substantial design discretion to the production stack. Second, a responsive interaction can provide a strong sense of control even when the builder cannot compare the resulting artifact with the wider output distribution. The process and survey analyses examine observable indicators relevant to these possibilities.

A growing HCI literature documents that ownership, authorship claims, and epistemic access can come apart in AI-mediated production. Users of text generators may report limited ownership of AI-generated text while still self-declaring as authors, the \emph{AI ghostwriter effect}, and felt ownership rises with influence over the output \citep{draxler2024}. Psychological ownership theory similarly links ownership feelings to exercised control, intimate knowing, and self-investment \citep{pierce2001}. Building on this literature, we propose \emph{authored ignorance} as a theoretical label for a narrower informational condition: a person can feel that they directed an artifact while lacking the comparison set needed to judge its position in an output distribution. Our survey measures perceived human control, satisfaction, and perceived quality, not perceived originality or awareness of genericity; the construct is therefore an interpretation to be tested directly in future work.

Two classical constructs supply vocabulary for the candidate process account. Norman's \emph{gulf of execution}, the distance between intention and available action, may in a chat interface partly take the form of translation from thought to prompt \citep{hutchins1985, norman1986}. The \emph{gulf of evaluation}, the distance between output and judgment, may widen when a finished-looking page invites a rapid holistic verdict while distinctiveness remains relational. Simon's satisficing describes acceptance of an option that clears an aspiration threshold rather than continued search \citep{simon1956}; related work shows that people sometimes follow algorithmic recommendations even when they could override them \citep{lane2026narrative, liel2025, wang2026power}. Section~\ref{sec:mechanism} reports indicators relevant to these ideas, including conditional transmission of verbalized intentions, verdict language, cycles without captured evaluation, and deferral. These indicators do not directly measure the gulfs or prove that one process variable caused another.

\subsection{Measuring Visual Similarity and Cohort Atypicality at Scale}

Concern about visual convergence on the web predates generative tools, as does computational analysis of web design. \citet{kumar2013} mined more than 100,000 rendered pages to characterize design demographics, and \citet{goree2021} combined computer-vision analysis of representative websites from 2003--2019 with designer interviews, reporting increased similarity after 2007 and attributing part of the trend to shared libraries, frameworks, and mobile standardization. This work provides historical context, not a direct baseline for the present cohort: its sampling frame, time period, rendering procedure, and similarity measure differ from ours. A comparable external corpus processed through the same pipeline remains necessary before the absolute similarity level observed here can be attributed to vibe coding or compared with prior production regimes.

Describing visual concentration at cohort scale requires a representation that can be applied consistently to thousands of pairs. Pixel-level metrics are poorly aligned with many perceptual relations, while expert ratings are costly at this scale. We use a self-supervised vision transformer from the DINO family as an embedding-based representation \citep{oquab2023, simeoni2025}. DINO features are sensitive to image content and structure, but our elongated full-page screenshots fall outside the natural-image settings in which these models are usually evaluated, and human pairwise calibration is still pending. We therefore report \emph{embedding-based visual similarity}, not a calibrated percentage of human-perceived similarity. Cosine similarities are multiplied by 100 for readability and called \emph{rescaled cosine-similarity index points}. A site's \emph{embedding-based cohort atypicality} is 100 minus its mean rescaled cosine similarity to the other 72 sites. Cohort-level summaries include nearest-neighbor similarity, descriptive cluster structure, occupancy-based diversity indices \citep{jost2006}, and the clustering-free order-2 Vendi score computed from the selected similarity kernel \citep{friedman2022vendi}. These quantities characterize the selected representation and reference set; they do not establish historical novelty, objective originality, or a literal number of designs. Adjacent traditions motivate the general strategy, from machine-learning models of product aesthetics \citep{burnap2023}, through measures of design prototypicality \citep{landwehr2011}, to distributional comparisons of generative text models \citep{pillutla2023}. Section~\ref{sec:method} details our implementation and robustness checks.

\section{Method}
\label{sec:method}

This section describes the deployment and its participants, the three layers of data it produced (published sites, builder surveys, and instrumented build sessions), how originality and diversity are measured, the baseline used to interpret the results, and the robustness and ethics of the design.

\subsection{Setting and Participants}
\label{sec:setting}

The participants were 84 graduate students enrolled in two sections of a required core course on the business applications of generative AI in a professional master's program in information systems at a large U.S. public university (the Foster School of Business at the University of Washington). The website task was a graded individual assignment that followed a three-hour lecture module on the best practices of vibe coding. The optional research component, which included a recording and a survey, was approved by our institutional review board and kept separate from grading. Students could decline to participate or participate only partially without penalty. Eighty-three students submitted the assignment. Seventy-seven submissions were publicly accessible websites at capture time. We excluded three sites unpublished for privacy, one submission that linked only to a recording, one internal company tool that was never publicly deployed, one software repository, and four interactive applications that were not comparable with scrollable promotional pages. The artifact analysis therefore covers 73 homepages. The separate reflection corpus contains 78 usable graded posts, All 83 students enter the reflection corpus described in Section~\ref{sec:reflections}.

Each student selected a real business that they could personally contact, such as local shops without websites, family businesses, non-governmental organizations (NGOs), restaurants, salons, and consultancies. Since no two students selected the same business, any convergence we find cannot come from convergence in the briefs. The cohort still shares a genre, assignment scope, or course context, which Section~\ref{sec:baseline} addresses.

The heterogeneity of the briefs can be stated more precisely. From the students' own business descriptions, confirmed independently against each homepage, the 73 sites span 31 kinds of business across nine broad groups (Table~\ref{tab:sectors}), from restaurants, taco trucks, and a speakeasy bar to wedding photographers, real-estate agencies, a Ugandan consultancy, a Saudi industrial contractor, a school in India, and a community center. Measured with the same index we later apply to the designs, the effective number of distinct business domains is 21.6 (inverse-Simpson over the 31 types; 25.7 by the Shannon equivalent).

\begin{table}
\caption{The 73 analyzed businesses by sector. Types and counts come from the students' own business descriptions, verified against each site's homepage.}
\label{tab:sectors}
\rev{\begin{tabular}{lrl}
\toprule
Sector & $n$ & Examples \\
\midrule
Food and drink & 15 & Cantonese bistro, taco trucks, coffee houses, speakeasy bar \\
Professional and trade services & 13 & real-estate agencies, dry cleaner, industrial contractor, import-export firm \\
Technology, apps, and media & 10 & IT consultancies, nightlife guide, career podcast \\
Creative services & 10 & wedding photography, risograph print studio, mandala artist, custom furniture \\
Health, wellness, and fitness & 9 & yoga studios, weightlifting gym, sauna and cold plunge, veterinary clinic \\
Retail and e-commerce & 6 & fabric shop, seafood market, plant nursery, stationery brand \\
Beauty and personal care & 4 & nail studios, esthetics studio, barbershop \\
Education and community & 4 & tutoring centers, a school, a community center \\
Travel and hospitality & 2 & short-term rental, travel guide \\
\bottomrule
\end{tabular}}
\end{table}

\subsection{Task and Tool}\label{sec:task}

Students built a promotional website within a defined scope: a homepage with a clear value proposition, business information, products or services, calls to action, functional navigation, mobile responsiveness, The website was tested on at least two devices (a computer and a phone) and two browsers. While the assignment strongly recommended Lovable, it permitted any AI-assisted tool. We analyzed the full published corpus, regardless of which tool students used. The grading rubric explicitly rewarded "creative and original design that avoids generic AI aesthetics." Students presented their website to at least one business stakeholder and one prospective customer. They documented the feedback and made changes based on it. Therefore, every site in our corpus was built for and shown to a real audience. Appendix~\ref{app:stakeholders} summarizes what that audience asked for and what became of the sites.

\subsection{Data: One Deployment, Three Nested Layers}
\label{sec:data}

The deployment yields three data layers, nested by construction, since every layer's participants are a subset of the layer above. Table~\ref{tab:layers} summarizes the layers, the research question each answers, and the measures each contributes.

\begin{table}
  \caption{The three nested data layers.}
  \label{tab:layers}
  \small
  \begin{tabular}{@{}lp{0.16\linewidth}p{0.30\linewidth}p{0.30\linewidth}l@{}}
    \toprule
    Layer & $N$ & Data & Primary measures & Answers \\
    \midrule
    Artifact & 73 & Full-page screenshots of every published site (1280-px viewport) & DINOv3 embeddings; pairwise similarity; per-site originality; archetypes; diversity indices & RQ1 \\
    Perception & 58 (57 for the authorship item) & Post-task survey; reflective blog post & Felt authorship; satisfaction; perceived quality; information foraging & RQ3 \\
    Process & 9 & Screen recordings with concurrent think-aloud, coded at one-second resolution & Behavior states; prompt cycles; spoken verdicts; translation loss; agency language & RQ2 \\
    \bottomrule
  \end{tabular}
\end{table}

\textbf{Artifact layer ($n = 73$).} Every published site was rendered to full-page screenshots at a standardized 1280-pixel viewport, homepage plus secondary pages. This layer answers RQ1.

\textbf{Perception layer ($n = 58$).} A post-task questionnaire measured felt authorship (``who drove, me or the AI,'' 0--100), satisfaction with the AI, perceived output quality, information foraging, prior experience, and demographics on validated scales; 58 of 73 sites join to a complete survey response (agency item: $n = 57$). A 500-word minimum reflective blog post, including a self-estimate of the AI-generated versus self-directed share of the site, supplements the survey. This layer answers RQ3.

\begin{figure}
  \centering
  \includegraphics[width=\linewidth]{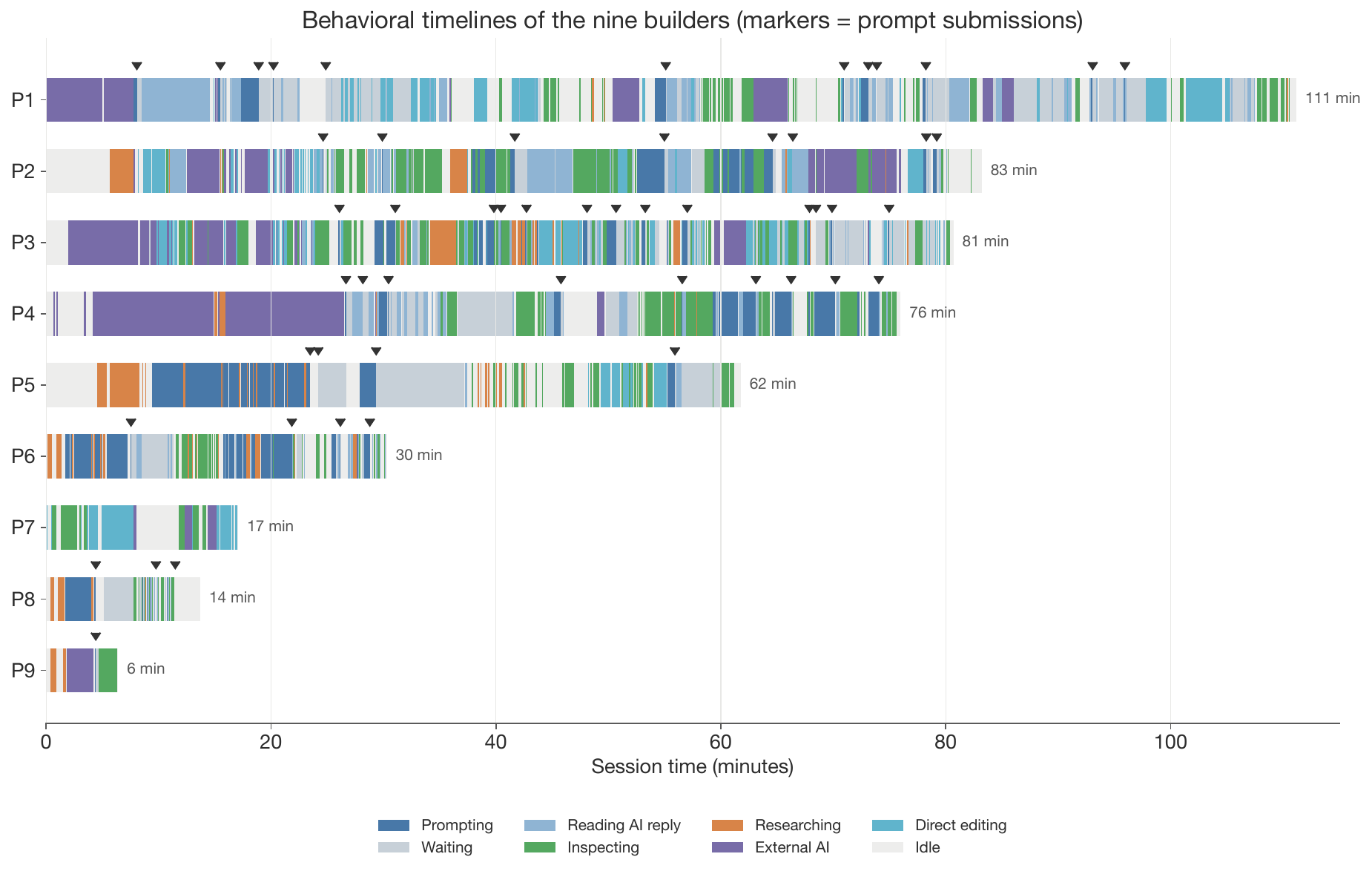}
  \caption{Behavioral timelines of the nine process-layer builders on a real-time axis, longest to shortest; triangles mark the 54 prompt submissions. One builder's opening block is a 25-minute prompt-drafting phase in an external AI; the hand-builder has no submission markers at all. Builders are ordered longest to shortest session.}
  \Description{Nine horizontal timelines showing color-coded behavioral states over real session time, with triangles marking prompt submissions.}
  \label{fig:timelines}
\end{figure}

\textbf{Process layer ($n = 9$).} Students optionally recorded their first build session via screen capture with a concurrent think-aloud protocol \citep{ericsson1993}, verbalizing prompts, reactions, and decisions as they worked. The recording corpus originally contained 24 sessions. A screening audit removed nine recordings that came from an earlier cohort performing a different task, and a second screen removed six that did not capture live building (finished-site walkthroughs whose extracted prompts were chat-history scrollback, a near-empty clip, and a session whose interaction occurred off camera). The analysis sample is therefore nine builders (sessions of 6--111 minutes; Figure~\ref{fig:timelines}). Seven drive the build through the platform chat, so the generate-and-check loop is on camera end to end; two are retained as informative contrast cases and flagged wherever relevant: one builds by hand in an editor and never submits a chat prompt on camera, and one is a six-minute fragment dominated by drafting in an external AI. This layer answers RQ2.

\subsection{Measuring Similarity, Originality, and Diversity}

Each homepage screenshot was embedded with DINOv3 \citep{simeoni2025} using native-resolution inference, in which the image is resized so its shortest edge is 224 pixels with aspect ratio preserved, and for the three of 73 pages whose full height exceeds a 10:1 aspect ratio we embed the top 10:1 region at native scale rather than compressing the whole page vertically\footnote{These three pages repeat the same design down their length (verified visually), so the top region carries the visual signal; the crop preserves true proportions.}, patch-aligned, and passed through the model in a single forward pass, yielding one CLS vector per page. This preserves the elongated geometry of webpage screenshots. We L2-normalize embeddings and compute cosine similarity for all $73 \times 72 / 2 = 2,628$ pairs.

From the similarity matrix we derive: per-site \textbf{atypicality}, which we report as "originality" following common usage in the creativity literature (100 minus mean similarity to the other 72 sites); \textbf{nearest-neighbor similarity} (each builder's closest look-alike); \textbf{archetypes} via UMAP projection followed by HDBSCAN density clustering \citep{mcinnes2018, campello2013}. Because density clustering on 73 points is configuration-sensitive, we report a sensitivity grid over 60 UMAP--HDBSCAN configurations (Section~\ref{sec:archetypes}; Appendix~\ref{app:robustness}) and use the resulting partition descriptively; the \textbf{effective number of distinct designs} via effective-diversity indices over archetype occupancy (inverse-Simpson and its Shannon equivalent, treating unclustered sites as singletons; \citealp{jost2006}), cross-checked by a clustering-free spectral count computed directly from the similarity matrix \citep[the Vendi score;][]{friedman2022vendi}; and \textbf{intrinsic dimensionality} estimates as a check against overclaiming collapse. Appendix~\ref{app:robustness} reports the measurement-robustness checks for this pipeline: page-length invariance under fixed-depth crops, agreement across geometry-preserving variants, localization of the convergence within the page, clustering sensitivity, and cross-model probes.

\subsection{The Interpretive Baseline}
\label{sec:baseline}

The cohort shares more than a platform: a genre (promotional small-business websites), an assignment scope (Section~\ref{sec:method}), one instructor, shared lectures and exemplars, and one rubric. The genre is convergent under any production regime, so the absolute similarity level reported in Section~\ref{sec:results1} requires an external anchor before it can be read as a property of the production stack. Therefore, we anchor our interpretation within the structure of the cohort (archetype concentration, nearest-neighbor statistics, and the originality spectrum) and describe the absolute percentages. The next step of this project is the external anchor, and the counterfactual is not one thing. A large share of real small-business sites are products of homogenizing production stacks, such as template builders and site wizards. The planned comparison corpus is therefore stratified, with same-sector samples of verified bespoke or agency-designed sites drawn from the pre-generative period via archival captures, template-builder sites, and the vibe-coded cohort, all processed through the pipeline of Section~\ref{sec:method} unchanged and reported with the same statistics per stratum. That design locates vibe coding on a convergence spectrum of production regimes and avoids testing it against an undefined baseline. Two further shared-context channels carry the same status, namely instructional convergence (shared exemplars and rubric language; Section~\ref{sec:process} documents rubric text being pasted into prompts, making the assignment itself part of the shared input) and peer visibility during the build window, which the cohort design cannot exclude.

\subsection{Process Instrumentation and Coding}
\label{sec:process}

All process layers share one clock at one-second resolution. The \emph{behavior} stream codes every second into eight states (Prompting, Waiting, ReadingAIOutput, Inspecting, Researching, ExternalAI, DirectEditing, Idle), extending the four-stage vibe-coding workflow of \citet{li2026} to second-level granularity. The \emph{think-aloud} stream is the time-aligned verbatim transcript with speech-onset snapping. The \emph{prompt} stream is a multi-pass extraction of prompts visible in the chat panel. First-pass coding of all three streams was performed by a frontier multimodal model (Gemini 2.5 Pro; temperature 0, JSON-constrained output, instructed to code conservatively), following emerging practice in LLM-assisted deductive coding \citep{xiao2023}; we state the validation status of every derived quantity below and in the results.

From the per-second codes we derive four master tables. A prompting \emph{bout} merges Prompting episodes separated by fewer than 30 seconds and counts as a \emph{submission} only if Waiting or ReadingAIOutput follows within 90 seconds (removing abandoned typing); a \emph{cycle} runs from one submission to the next; a \emph{generation interval} merges Waiting episodes across gaps of up to 90 seconds, approximating the span in which a generation was plausibly running. The 54-cycle set spans the eight builders who submitted at least one on-camera prompt, including the contrast cases (the visual-editor builder contributes four cycles and the external-AI fragment one; the hand-builder contributes none), and cycle-level analyses flag the contrast cases wherever they matter. Table~\ref{tab:units} lists the units. The statistical leverage comes from working below the person level: claims are made at the cycle, episode, or utterance level, with the nine sessions treated as replications. A pattern that persists session after session supports analytic generalization even though the person-level $N$ is small; we draw no inferences between-persons.

\begin{table}
  \caption{Process-layer units of analysis.}
  \label{tab:units}
  \begin{tabular}{lrl}
    \toprule
    Unit & $N$ & Used by \\
    \midrule
    Coded seconds & 28{,}818 (8.0 h) & all sequence analyses \\
    Behavior episodes (uninterrupted spells) & 1{,}652 & survival, motifs, flows \\
    Prompt cycles (submission to next) & 54 & event studies, waiting economics, verdicts \\
    Generation intervals (merged Waiting clusters) & 54 & waiting economics \\
    Speech utterances & 3{,}542 & all language analyses \\
    Extracted prompt texts & 25 & prompt analyses \\
    Spoken verdicts (evaluative utterances) & 440 & verdict lexicon \\
    \bottomrule
  \end{tabular}
\end{table}

Every utterance was optionally labeled \emph{eval} (polarity and the criterion invoked: aesthetic, functional, business fit, originality, other, with verbatim evaluative terms), \emph{breakdown} (something reported broken or wrong), \emph{deferral} (a decision explicitly ceded or settled; subtypes aesthetic, satisfice, fatigue), \emph{discovery} (a tool capability or limitation realized), and \emph{agency} (creation work credited to the speaker or to the AI through a making verb). Prompts were coded for speech acts (content supply, build request, aesthetic directive, constraint, gradient modification, bug report, meta), target surface (the platform versus an external drafting AI), gradient adjectives (the $X$ in ``make it more $X$''), and suspected pasted assignment text (LLM judgment plus $n$-gram screening; a definitive verbatim test was not possible because the original assignment document is unavailable).

\emph{Loop conformance} reduces each session to its sequence of active states and scores the share of transitions consistent with the canonical generate-and-check loop (Prompting $\to$ Waiting $\to$ ReadingAIOutput $\to$ Inspecting $\to$ Prompting), tested against a within-person null of 1{,}000 shuffled sequences; \emph{motif mining} counts frequent 3--5-state subsequences against a per-person first-order Markov null. \emph{Peri-event histograms} align all 54 cycles at prompt submission and at generation end; because the generation end is defined behaviorally (the last second of the Waiting cluster), second-level latencies around that event carry construction noise, which we note where relevant. \emph{Waiting economics} decomposes each generation interval second by second into dead waiting, productive fill, peeking at the preview, and idle. \emph{Translation loss} compares, for every cycle with usable prompt text, the atomic intentions voiced in the 240 seconds before submission against the atomic requirement units of the typed prompt, marking each spoken intent transmitted or dropped; because think-aloud does not capture all intention, the transmission rate is conditional on verbalized intent, and we phrase results accordingly. \emph{Breakdown--repair tracing} follows each spoken breakdown forward and logs the ordered repair actions read off the behavior stream (reprompt, direct edit, research, external AI, or none) and whether a positive verdict follows within the window. \emph{Agency} is measured twice: a rule-based count of first-person versus AI-subject constructions with creation verbs (narration of action), and the model's label of who is credited with the making (attribution).

The behavioral codes are the result of an initial AI analysis. In a pre-study, human coders reached an agreement on a related scheme $\kappa = 0.74$. A stratified 54-utterance validation sample, a full review of the 92 breakdown--repair ladders, and a spot check of translation-loss pairs are in progress, and utterance- and prompt-level quantities are read as first-pass estimates until they complete. Because spot checks revealed that the discovery label was triggering too often, discovery counts are not reported. Prompt-extraction timestamps were unreliable when participants scrolled through the chat history because old prompts would surface in bursts. Therefore, every extracted prompt was validated against the behavior stream. Raw timestamps are only used where a time-alignment flag permits. Prompt texts and their chronological order are used for all text analyses. The supplementary codebooks, coding prompts, and validation materials accompany the archival version.

\subsection{Ethical Considerations}

The deployment sits inside a graded course, and Section~\ref{sec:setting} describes how the design separates research participation from grading. Beyond that separation, all analyses are reported at the cohort or sub-person level, artifact screenshots are shown with identifying labels anonymized, and because participants deployed their sites publicly, we report distributional statistics and never link a named site to its originality score or process behavior. Verbatim quotations from the think-aloud recordings and from the written reflections are limited to short fragments and are quoted without names, identified at most by the kind of business the student built for.

\section{Results I: Within-Cohort Visual Concentration}
\label{sec:results1}

\subsection{Pairwise and Nearest-Neighbor Similarity}

\begin{figure*}
  \centering

  \includegraphics[width=\linewidth]{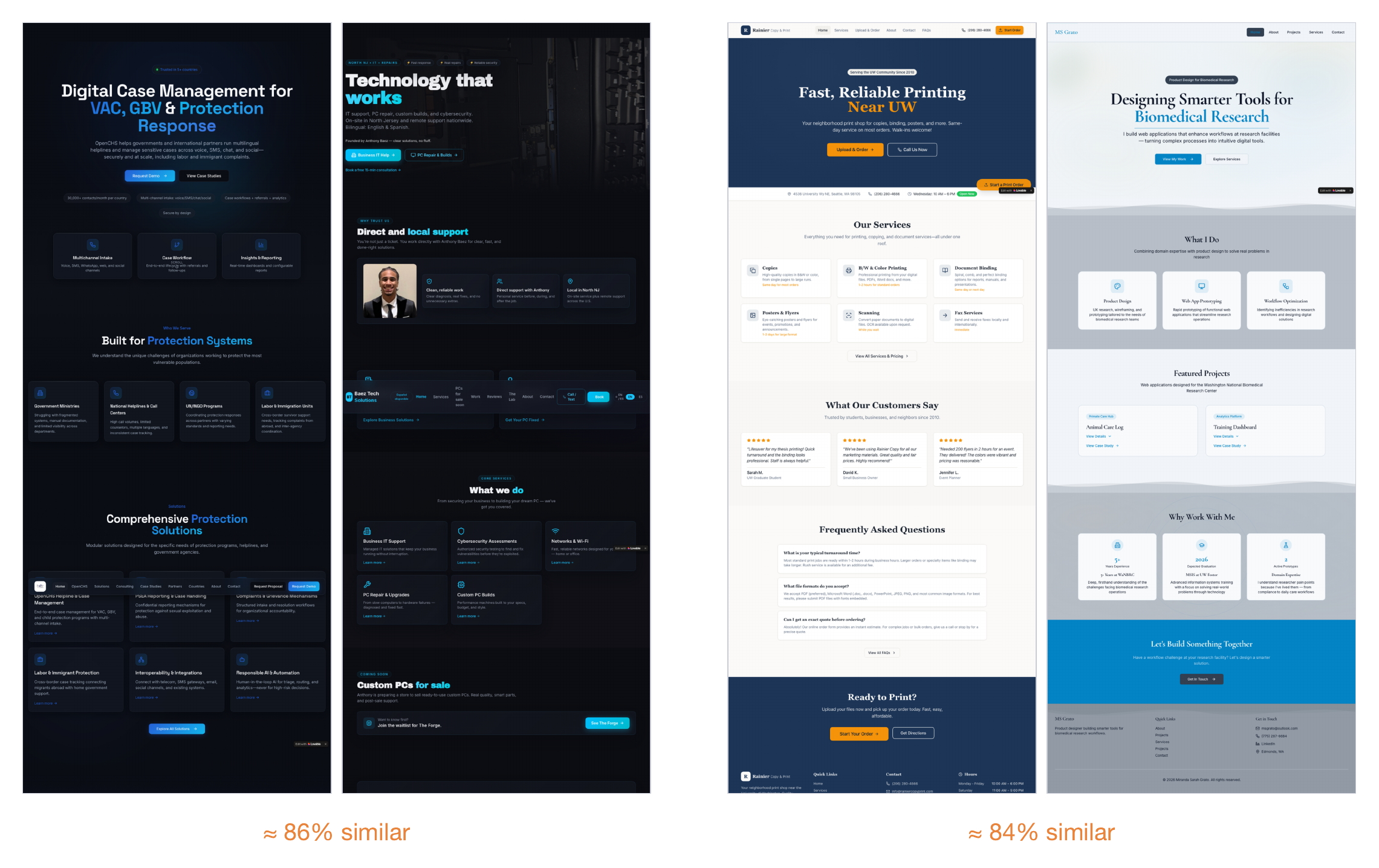}
  \caption{The two highest-similarity pairs under the DINOv3 full-page representation, each joining distinct businesses built by different students and shown to a common scroll depth. The pairs score approximately 86 and 84 rescaled cosine-similarity index points. These are embedding scores, not calibrated percentages of human-perceived similarity. Screenshots must be de-identified before public release.}
\Description{Four full-page website screenshots side by side in two pairs; within each pair, the two unrelated businesses' pages share the same layout and section order from top to bottom.}
  \label{fig:twins}
\end{figure*}

\begin{figure}
  \centering
  \includegraphics[width=\linewidth]{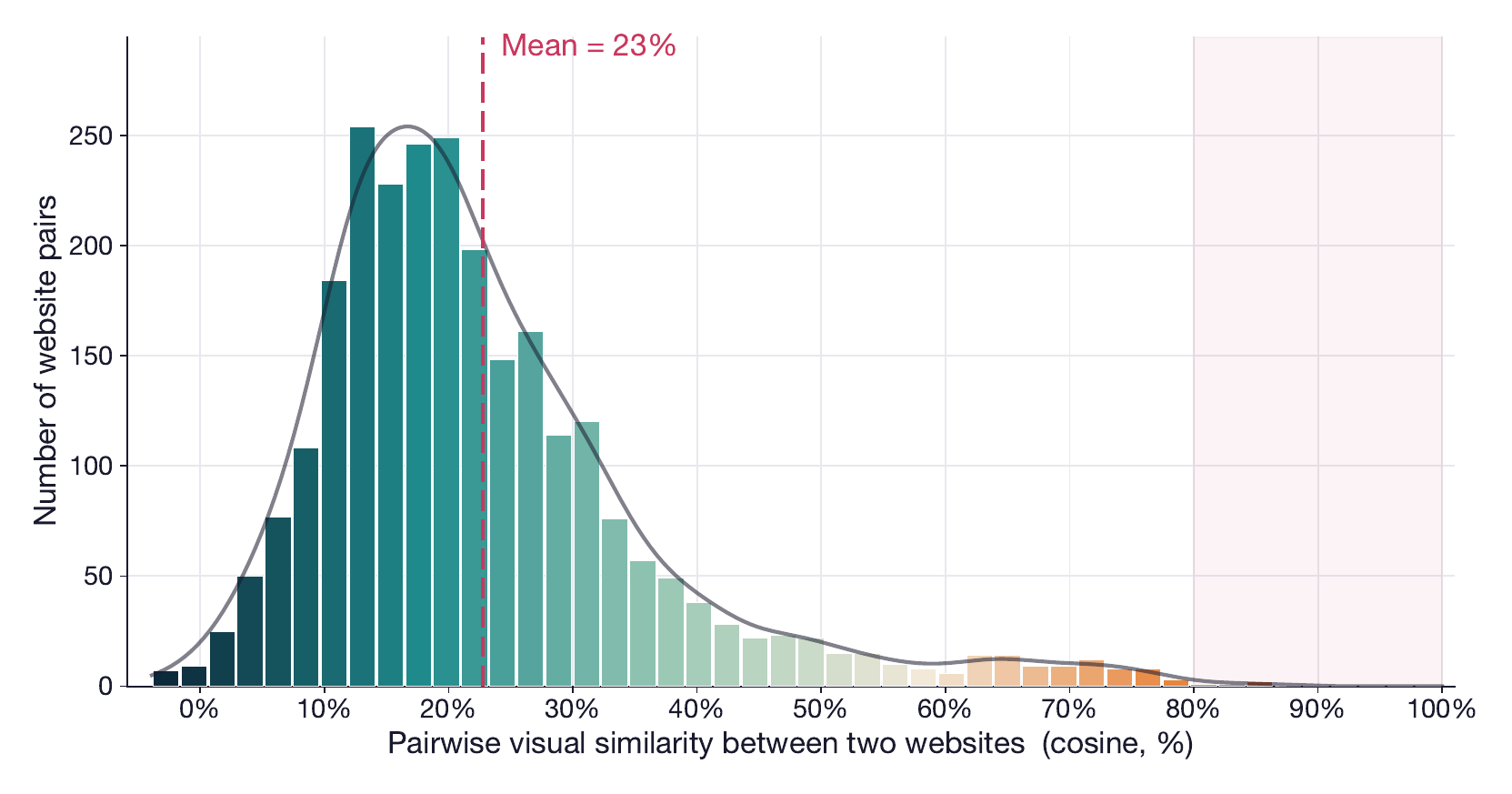}
  \caption{Distribution of all 2,628 pairwise DINOv3 cosine-similarity scores after multiplication by 100. The mean is 22.8 index points, with a right tail of high-scoring pairs. The scale is not calibrated as a percentage of perceptual similarity.}
  \Description{Histogram of pairwise similarity scores, right-skewed with mean at 22 percent and a tail extending past 80 percent.}
  \label{fig:spectrum}
\end{figure}

Across all 2,628 pairs, mean rescaled cosine similarity is 22.8 index points (median 19.7; Figure~\ref{fig:spectrum}). The distribution is right-skewed: 38 pairs exceed 70 index points and four exceed 80. These thresholds are descriptive properties of the DINOv3 representation rather than human-calibrated cutoffs. The highest-scoring pair consists of two distinct IT consultancies built by different students and scores 86 index points (Figure~\ref{fig:twins}), with a similar dark hero, gradient accents, and section rhythm. We did not observe or measure all peer interactions, so we do not claim that the builders never encountered one another's work.

\begin{figure}
  \centering
  \includegraphics[width=\linewidth]{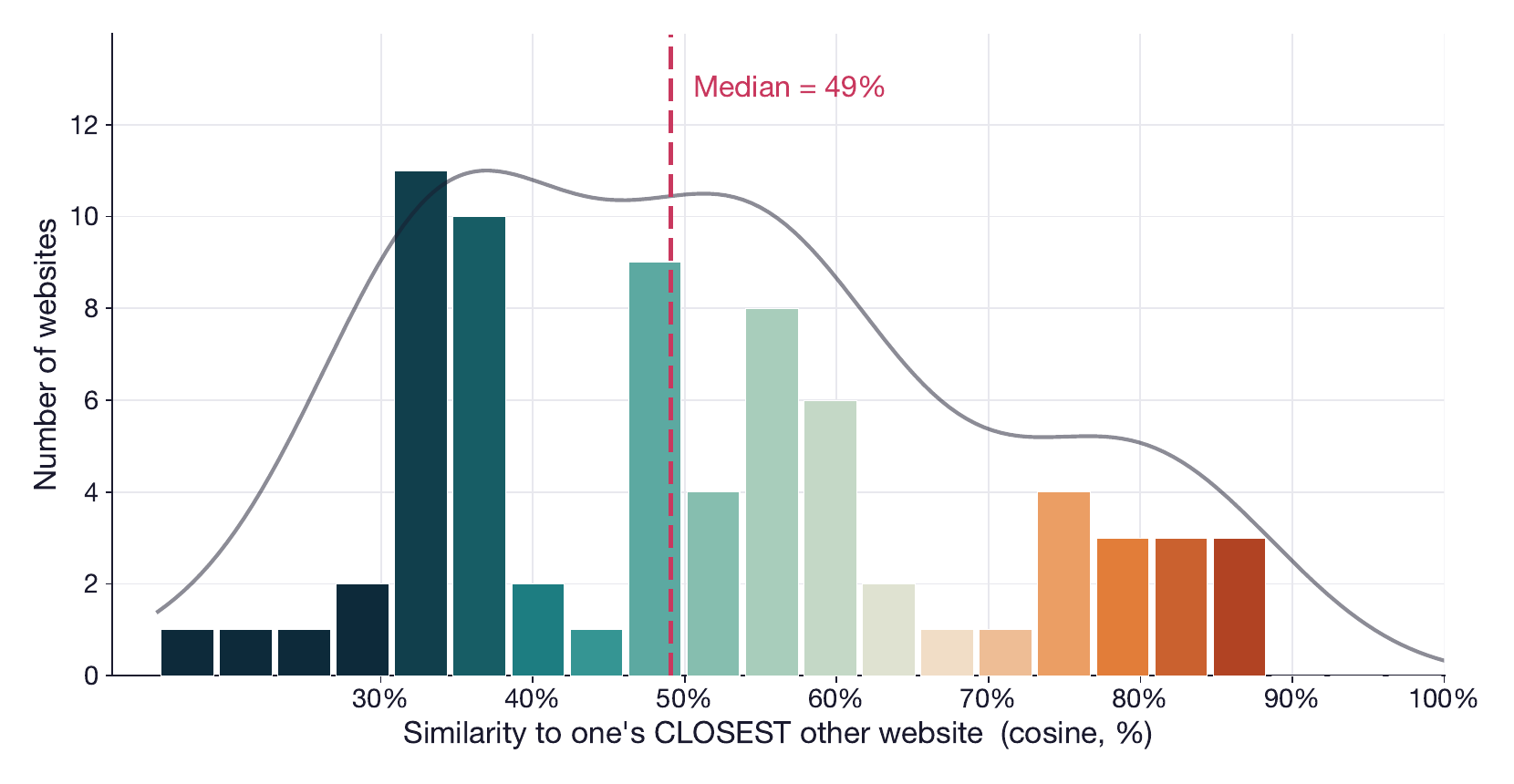}
  \caption{Each site's rescaled cosine similarity to its closest other homepage in the cohort. The median nearest-neighbor score is 49 index points, compared with a mean of 22.8 across all pairs.}
  \Description{Distribution of nearest-neighbor similarities centered around 53 percent.}
  \label{fig:doppelganger}
\end{figure}

Similarity is locally concentrated. Each site's closest neighbor among the other 72 has a median score of 49 index points, compared with 22.8 for the average pair (Figure~\ref{fig:doppelganger}). Fourteen of 73 sites (19\%) have a nearest-neighbor score above 70. The sites were built for distinct business targets, but shared genre and assignment conventions remain plausible contributors to these close visual counterparts.

\subsection{A Representative Six-Cluster Description}
\label{sec:archetypes}

\begin{figure*}
  \centering
  \includegraphics[width=\linewidth]{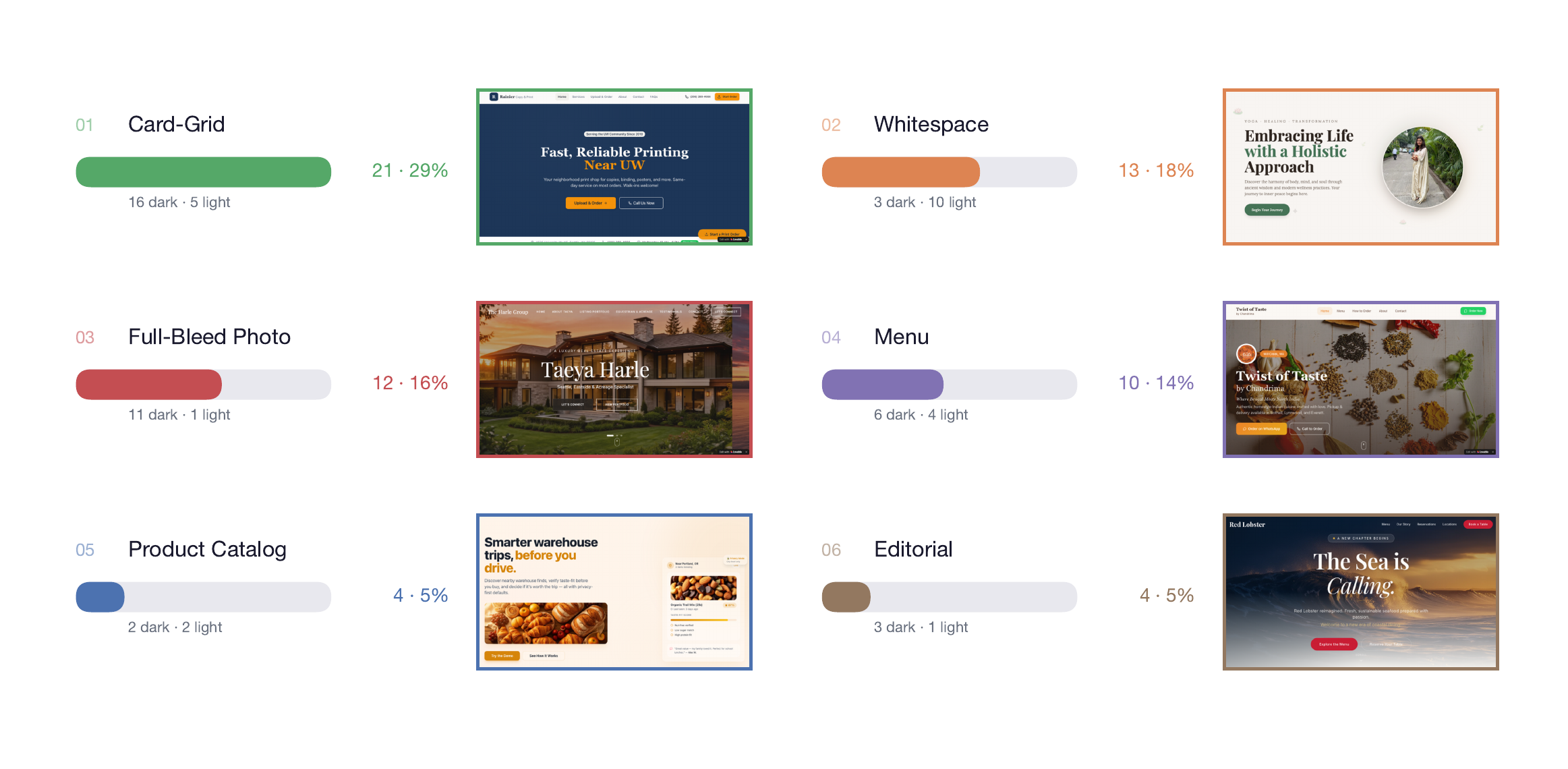}
  \caption{One illustrative UMAP--HDBSCAN specification yields six interpretable clusters and assigns 64 of 73 sites (88\%); nine points are density noise. Across 60 specifications, the number of clusters ranges from three to twelve, so the six labels are post hoc descriptive summaries rather than stable natural kinds. The displayed screenshot is the highest mean within-cluster cosine member and must be de-identified before public release.}
  \label{fig:housestyles}
\end{figure*}

\begin{figure*}
  \centering
  \includegraphics[width=\linewidth]{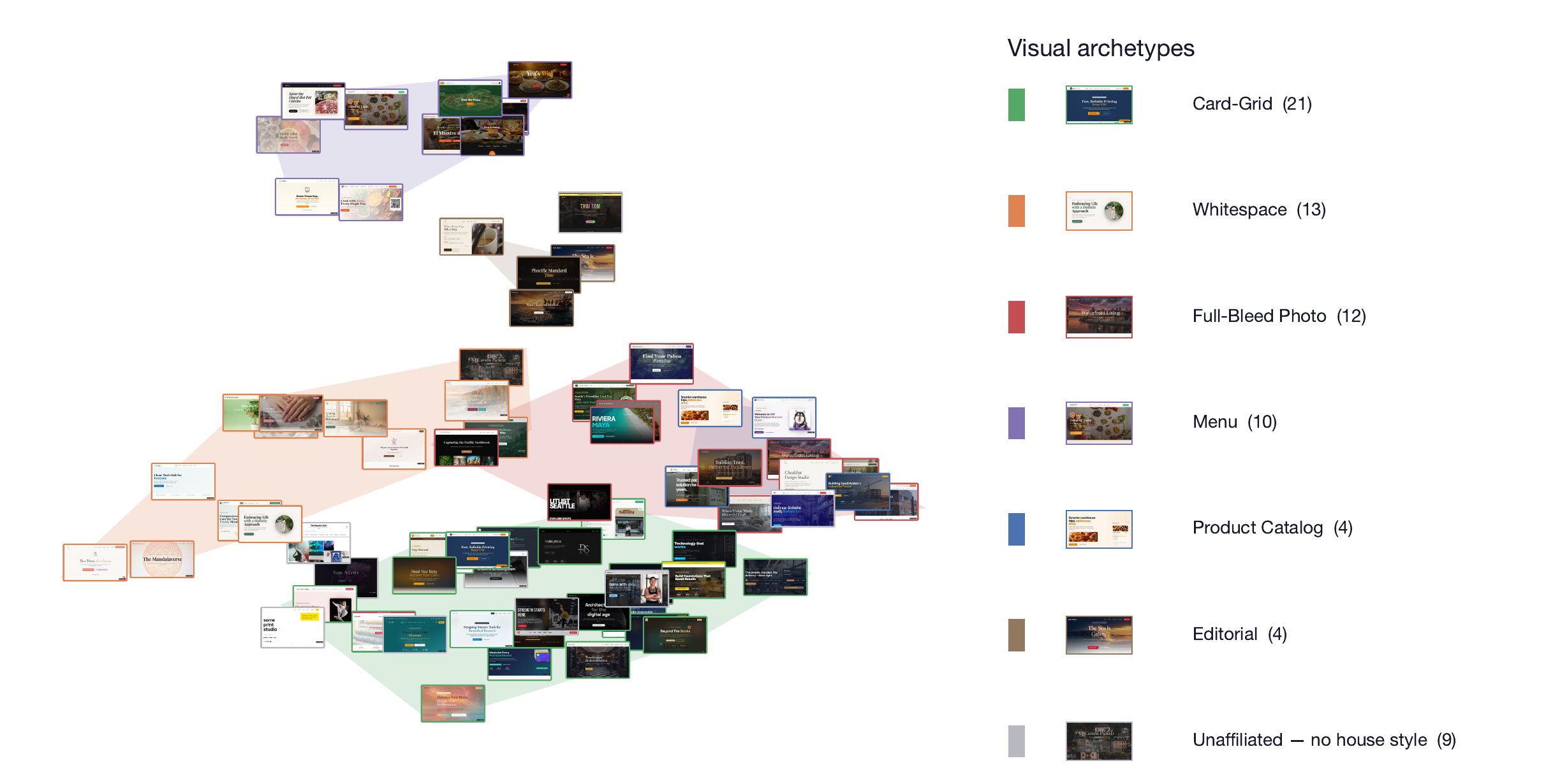}
  \caption{UMAP projection of the full-page DINOv3 embeddings under the illustrative six-cluster partition. The projection is a descriptive visualization; distances and cluster boundaries should not be interpreted as a unique map of an ``AI design space.'' Thumbnails must be de-identified before public release.}
  \Description{Two-dimensional scatter map where website thumbnails cluster into colored groups.}
  \label{fig:map}
\end{figure*}

Density clustering of the embedding space recovers six visual archetypes, which absorb 64 of 73 sites (88\%; Figure~\ref{fig:housestyles}). The count of six is the answer of our recorded clustering configuration (Appendix~\ref{app:robustness}) rather than a sharp fact about the cohort: across a sensitivity grid of 60 UMAP--HDBSCAN configurations, the archetype count ranges from 3 to 12 (median 7), so we use the six-archetype partition descriptively, as a legible summary of density structure whose quantitative content the clustering-free statistics below confirm. We name the six house styles by their structural signature rather than their palette. The \emph{Card-Grid} (a hero band over a grid of uniform feature cards) is the cohort's default at 21 sites (29\%), spanning businesses as unrelated as two IT firms, a barber, and a credit-card-rewards app; the others are the \emph{Whitespace} (airy, low-density, pastel), the \emph{Full-Bleed Photo} (an edge-to-edge photographic hero), the \emph{Menu} (a food hero over dish blocks), the \emph{Product Catalog} (a text hero over dense feature cards), and the \emph{Editorial} (an atmospheric photo with serif type). Palette is orthogonal to these layouts, splitting within each style rather than defining it: the Card-Grid is 16 dark and 5 light, and the most dark-dominant style is in fact the Full-Bleed Photo (11 of 12), so a ``dark'' look is a skin applied across layouts rather than an archetype of its own (Figure~\ref{fig:housestyles}; the hero-level layout~$\times$~palette structure is quantified in Appendix~\ref{app:robustness}). Nine sites fall outside every archetype (Figure~\ref{fig:map}). Unclustered status is a statement about local density: these sites were simply not absorbed into one of the six dense cores. They are no sparser than clustered sites (median nearest-neighbor similarity 55\%, versus 48\% for clustered sites) and no more original (mean originality 79\% versus 77\%, n.s.); the cohort's single most original site is itself clustered. We read them as unaffiliated, and reserve originality claims for the per-site spectrum of Section~\ref{sec:spectrum}.

Summarizing this collapse with diversity indices, the effective number of distinct designs is roughly six to twelve, depending on the estimator. The clustering-free spectral count computed directly from the similarity matrix, with no clustering step at all, gives 12.0 (order-2 Vendi score; Appendix~\ref{app:robustness}); occupancy-based indices over the six-archetype partition give 8.0 (Shannon-equivalent) and 6.0 (inverse-Simpson, which weights the largest archetype more heavily). The two families answer different questions --- the parameter-free Vendi count, which we read as primary, does not depend on the clustering choice, while the occupancy indices track the coarser partition --- but on any of them the design space is far narrower than the input. The input space was roughly twice as diverse in parameter-free terms (21.6 inverse-Simpson across the 31 business types, against the 12.0 Vendi count for the designs). Seventy-three independent design problems, processed using a single tool, yielded roughly a dozen designs.

The convergence is not confined to layout; it recurs on the other axes a viewer registers at a glance. Color palettes reduce to a shared neutral canvas plus a narrow band of accent hues---about three effective palettes, with warm/earth tones and trust-blue dominating---and typography reduces to a common type system, a serif display heading (most often Playfair Display or Cormorant Garamond, together over half the cohort) set over an Inter or DM-Sans body (Inter alone on half the sites). We quantify both the palette and the type convergence in Appendix~\ref{app:robustness} (Figures~\ref{fig:palettewheel}--\ref{fig:fonts}).

The sector data of Table~\ref{tab:sectors} lets us ask how much of this structure the briefs themselves explain. The answer is: some, but not much. The association between a site's business sector and its archetype is real but modest (adjusted mutual information $0.20$ between the nine sector groups and the archetype assignment, $0.26$ for the 31 fine-grained types; both exceed a permutation null at $p < .001$). Some archetypes are visibly genre-flavored, most clearly the warm food hero that collects restaurants, so the production stack's defaults are conditioned on genre rather than uniform across it. But knowing a site's sector removes only about a fifth of the uncertainty about which house style it received, and every large archetype spans several sectors: the biggest gathers two IT firms, a barber shop, and a credit-card-rewards app under the same dark hero. Sector is at most weakly associated with how original a site is (Kruskal--Wallis $p = .064$ across sector groups, not significant): retail and creative-services sites tend highest (median originality 82\% and 82\%), while technology and health-and-wellness sites tend lowest (70\% and 72\%). It is notable that the businesses closest to the model's most common training material, technology companies and apps, received the most interchangeable designs.


\subsection{The Originality Spectrum}\label{sec:spectrum}

\begin{figure}
  \centering
  \includegraphics[width=\linewidth]{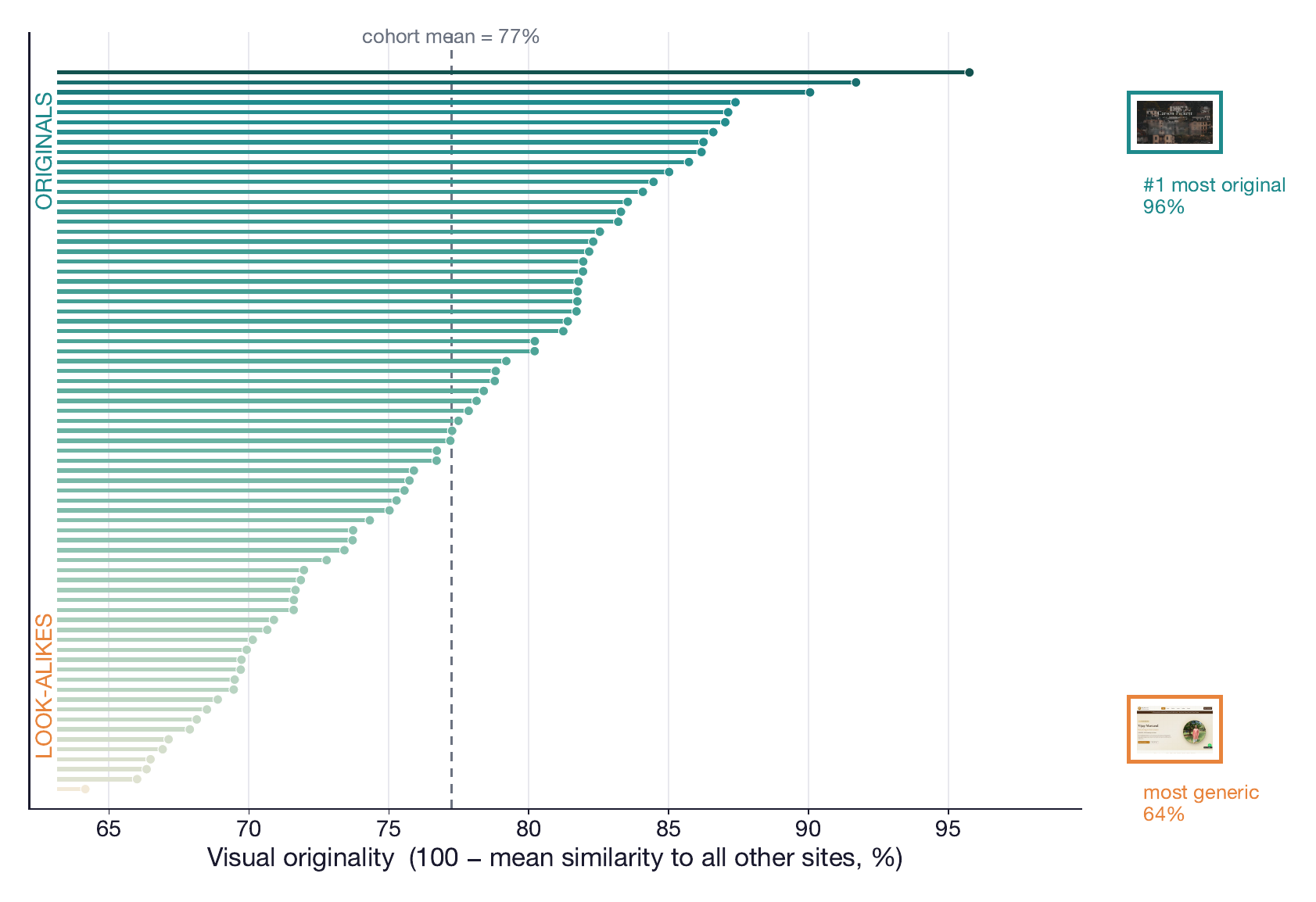}
  \caption{All 73 sites ranked by visual originality, showing a few originals and a long tail of look-alikes (100 minus mean similarity to the rest of the cohort). }
  \Description{Ranked lollipop chart of originality scores from most generic to most original.}
  \label{fig:ranking}
\end{figure}

\begin{figure*}
  \centering
  \includegraphics[width=\linewidth]{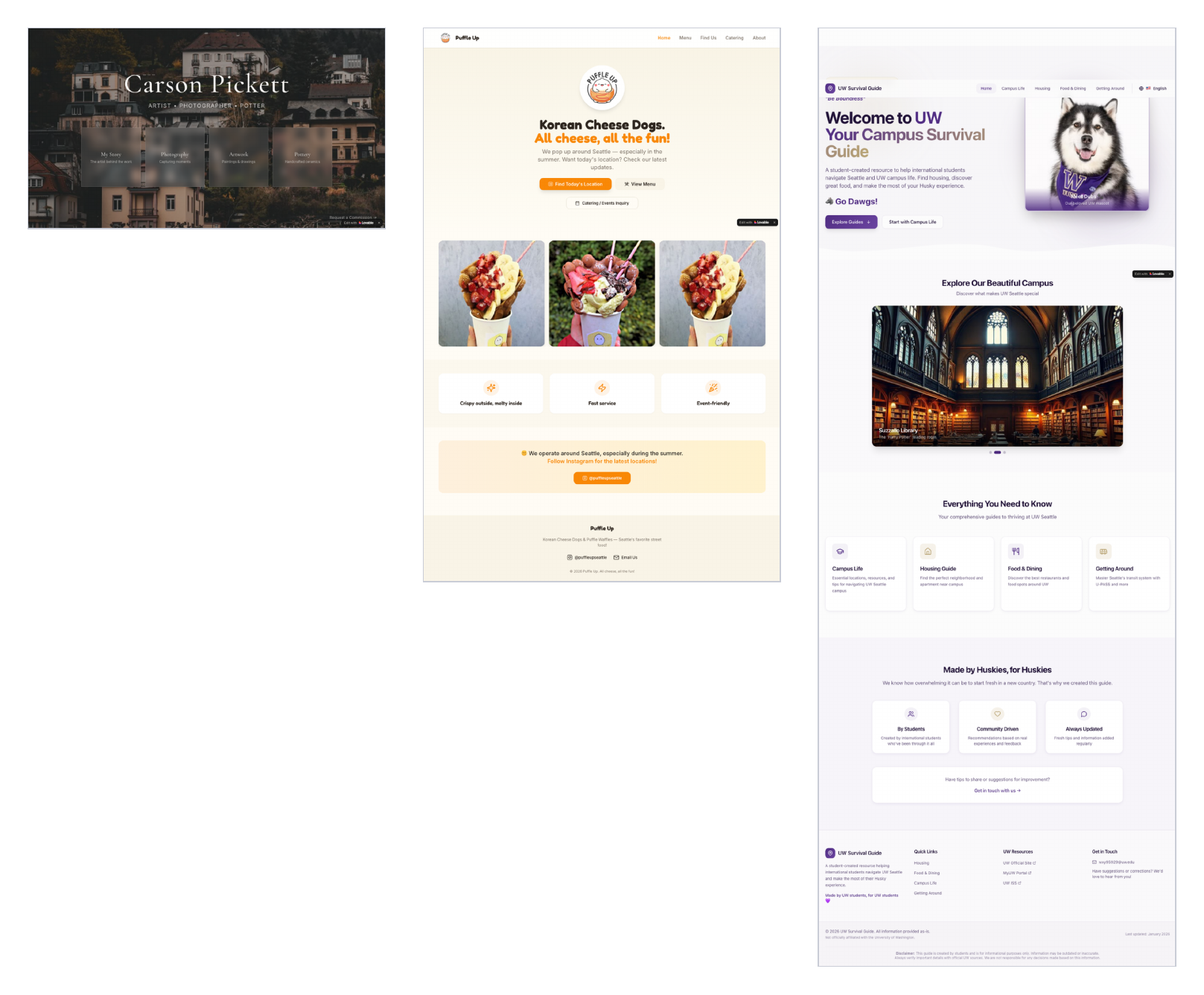}
  \caption{Three of the cohort's most original sites, each departing from the recurring house
  styles in a different direction: a full-bleed editorial photograph with a serif
  wordmark and glassmorphic cards (left); a dark, almost UI-less cinematic hero
  (centre) to create an immersive experience; and a playful mascot brand built from rounded forms
  (right) and with actual pictures. Full pages shown as captured.}
  \label{fig:originals3}
\end{figure*}

Per-site originality averages 77\% and spans a wide range (Figure~\ref{fig:ranking}). The cohort's most original site shares just 4\% mean similarity with the field, a genuinely idiosyncratic design, while the most generic shares 36\%, sitting deep inside one of the house styles. Originality, in other words, was achievable, and some builders escaped. Figure~\ref{fig:originals3} shows three of the cohort's most original sites at full page, each departing from the recurring house styles in a different direction.


\subsection{From the Artifacts to the Process}

These findings provide a profound yet structured answer to RQ1 under maximal input diversity and an explicit incentive to diverge. This result clarifies RQ2. Identical outputs from divergent inputs imply the existence of something shared upstream of the artifacts, and the embedding space can only point to it. The recurring archetypes resemble the production stack's typical styles. However, how those styles survive seventy-three different builders with seventy-three different briefs is not evident in the screenshots. This is visible in the build sessions, which are examined in the next section. Nine sessions were instrumented to one-second resolution and show the candidate mechanism in operation.

\section{Results II: The Mechanism of One Script, Thin Input, and Shallow Acceptance}
\label{sec:mechanism}

If the production stack's defaults capture the artifact, the capture must happen somewhere in the loop. The process data lets us watch for it in three places, namely the structure of the loop itself (is there one way of building, or many?), the keyboard (how much does the human actually specify?), and the moment of judgment (what does it take for an output to be accepted?). All quantities below derive from the instrumentation of Section~\ref{sec:process}, which states their coding provenance; claims are made at the cycle, episode, and utterance level with the nine sessions as replications.

\subsection{One Script}
\label{sec:onescript}

We check whether the building process itself is uniform across builders. This matters because if a shared default explains why the outputs look similar, that explanation only holds if builders actually use the tool in the same way. We test this in two steps.

The first step checks whether each session has any organized structure, rather than random behavior. For each builder, we measure how often their actual sequence of behavioral states matches the canonical generate-and-check loop (Prompting $\to$ Waiting $\to$ Reading $\to$ Inspecting $\to$ Prompting), and we compare that rate to what a randomly shuffled version of the same session would produce. For all seven chat-driven builders, the actual session matches the canonical loop far more often than the shuffled version does (conformance 0.21--0.80; $z = +2.8$ to $+6.2$ across all seven). This is a low bar, because almost any task with distinct phases would pass it. So this result only shows that each builder's session is organized. It does not show that the nine builders are organized in the same way.

\begin{figure}
  \centering
  \includegraphics[width=\linewidth]{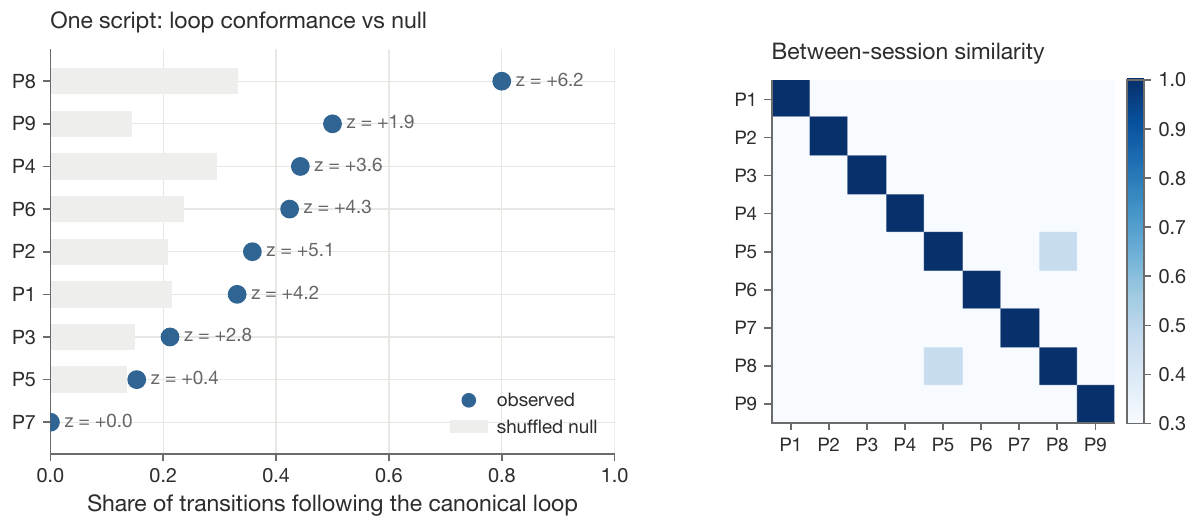}
  \caption{One script. Left: share of transitions following the canonical generate-and-check loop (dots) against a within-person shuffled null (bars); every chat-driven builder exceeds the null. Right: between-session similarity over normalized time; pointwise agreement is modest, so the shared object is the loop, not a schedule.}
  \Description{Left panel compares loop conformance per builder against shuffled null distributions; right panel shows a matrix of pairwise between-session similarity.}
  \label{fig:onescript}
\end{figure}

The second step checks that stronger claim directly. We use motif mining to look for short sequences of states that recur across builders more often than a per-person first-order Markov model would predict. This kind of model predicts each next state using only the current state, based on that person's own data. The strongest recurring sequence is Prompting $\to$ Waiting $\to$ Reading $\to$ Waiting. It occurs 21 times, about 1.89 times more often than the model predicts, and appears in at least six of the nine sessions. This sequence happens when a builder waits for the AI to generate a response, reads the response, and then waits again. Figure~\ref{fig:onescript} (right panel) shows how similar the sessions are to each other, which clarifies what we mean by "one script." When we line up all sessions on a common, normalized timeline and compare behavior at each matching point, the agreement between sessions is modest. So builders do not follow the same schedule. What they do share is the loop itself and its recurring motifs: the same cycle, run at different speeds and for different lengths of time.


The script also has a consistent internal shape, which we call the reflex arc. On average, a session begins with a brief period of orientation, reaches its peak rate of prompting in the first quarter, and spends the second half mostly on inspection. Each behavioral episode is short: a waiting episode lasts a median of eight seconds before attention shifts, and an inspecting episode lasts a median of nine seconds.

To see the reflex arc directly, we align all 54 cycles at two reference points: the moment a prompt is submitted, and the moment a generation ends.

Aligned at submission: the probability of prompting is highest right at submission, the probability of waiting spikes immediately afterward, the probability of reading is highest around 50 seconds later, and the median time from submission to the first look at the preview is 15 seconds.

\begin{figure}
  \centering
  \includegraphics[width=\linewidth]{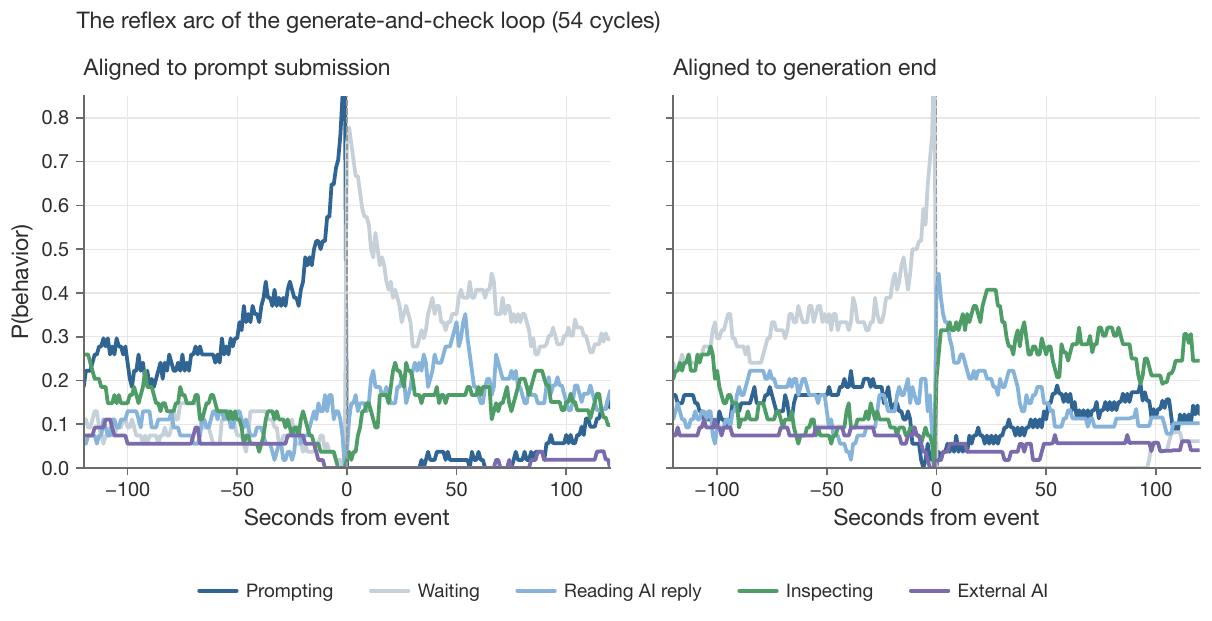}
  \caption{The reflex arc. Behavior probabilities around prompt submission (left) and generation end (right), pooled over 54 cycles. The $t=0$ alignment in the right panel is behaviorally defined, so its discontinuity is partly by construction; the decay afterward is not.}
  \Description{Two peri-event panels showing the probability of each behavioral state second by second around prompt submission and generation end.}
  \label{fig:eventstudy}
\end{figure}

Aligned at generation end: the probability of inspecting rises to 0.34 in the first thirty seconds after generation ends, then falls to 0.26 within two minutes. So most checking happens right after generation ends, and it does not last long (Figure~\ref{fig:eventstudy}).

\begin{figure}
  \centering
  \includegraphics[width=\linewidth]{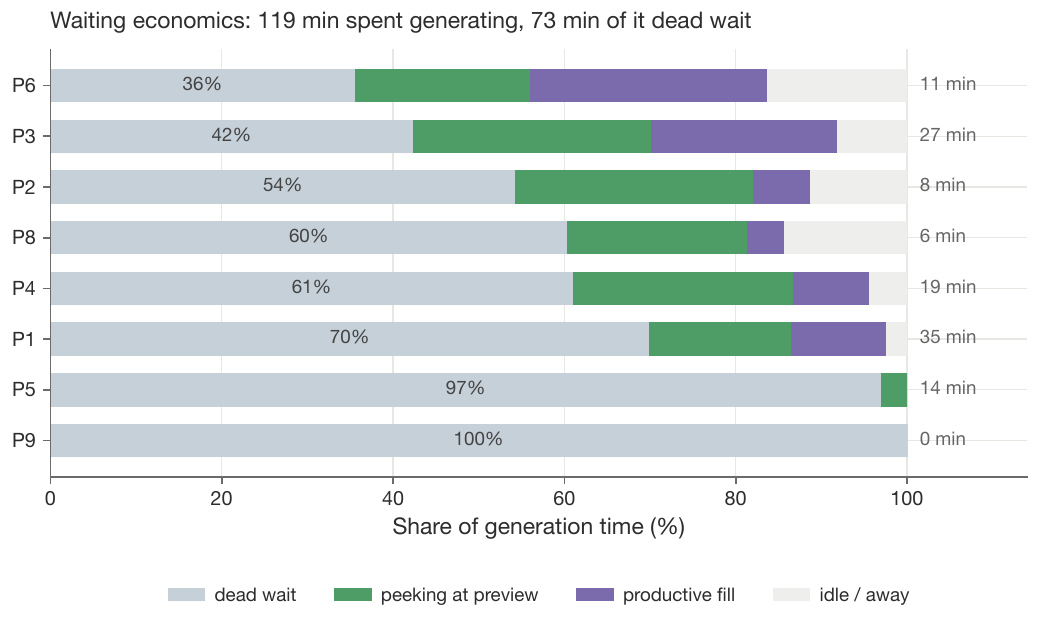}
  \caption{Waiting economics. Composition of generation time per builder, ordered by dead-wait share; the right margin gives each builder's total generation minutes. Across builders, 61\% of generation time is dead waiting.}
  \Description{Stacked horizontal bars per builder decomposing generation time into dead waiting, peeking, productive fill, and idle.}
  \label{fig:waiting}
\end{figure}

Most of the time in the loop is spent waiting for the tool. Generation intervals cover 119 minutes, a quarter of all recorded time. Within those intervals, pooled across all generation seconds, 61\% is dead waiting (73 minutes total, watching the spinner), 20\% is peeking at the preview, 13\% is productive fill, and 6\% is idle (Figure~\ref{fig:waiting}).

Builders learn to fill some of this time: productive fill rises from 9\% in the first half of a person's intervals to 17\% in the second half. But waiting is still the single largest activity in the loop, and builders mostly just watch it happen.

The most frequent behavioral transitions are between chat and preview (Chat to Preview: 80 episode transitions; Preview to Chat: 66), showing this is the main path builders follow during a session. The editor and the open web both feed back into chat.

\subsection{Thin and Thinning Human Input}
\label{sec:thininput}

\begin{figure}
  \centering
  \includegraphics[width=\linewidth]{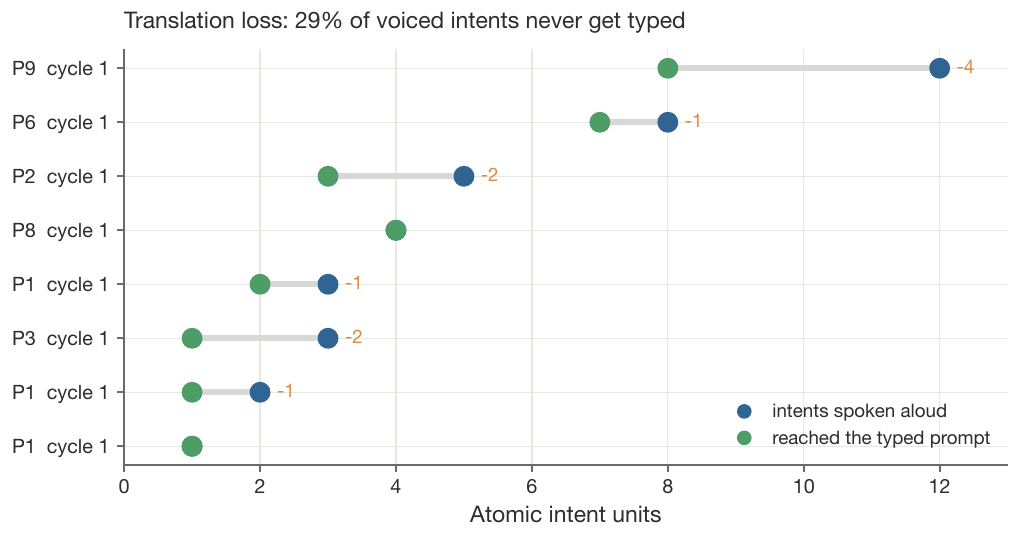}
  \caption{Translation loss. Spoken intents versus intents that reached the typed prompt, for each of the eight analyzable cycles; 29\% of voiced intentions (38 intents total), including explicit quality requirements, never reach the model. Illustrative given the small denominator; see Section~\ref{sec:thininput}.}
  \Description{Per-cycle comparison of the number of voiced intentions and the subset transmitted into the typed prompt.}
  \label{fig:translation}
\end{figure}

The second question is how much of the design the human actually specifies, because whatever goes unspecified is decided by the production stack's defaults. The answer is observable at the keystroke. Across the eight prompt cycles with a usable speech window, spanning six of the nine builders, builders voiced 38 atomic intents and typed 27 of them, so 71\% were transmitted and 29\% were dropped between mind and keyboard (Figure~\ref{fig:translation}). The denominator is small, and we treat the percentage as an illustration and draw no population estimate from it; the qualitative finding does not depend on it. The dropped intents are not noise; they include explicit quality requirements such as \emph{do not use any AI-generated images} and \emph{images should load immediately}. In the other direction, 39\% of typed prompt units were never voiced beforehand, much of it pasted boilerplate. Intentions are richer than prompts, and whatever does not survive the typing step is decided by the production stack's defaults. Two scope conditions apply. The transmission rate conditions on verbalized intent; think-aloud does not exhaust intention. And speech is cheap in any medium: without a cross-medium anchor (designers briefing human developers, users completing a structured intake), we cannot say whether 29\% is high; the loss is real, and what is lost defaults to the stack. A comparison condition is natural future work (Section~\ref{sec:limits}).

\begin{figure}
  \centering
  \includegraphics[width=\linewidth]{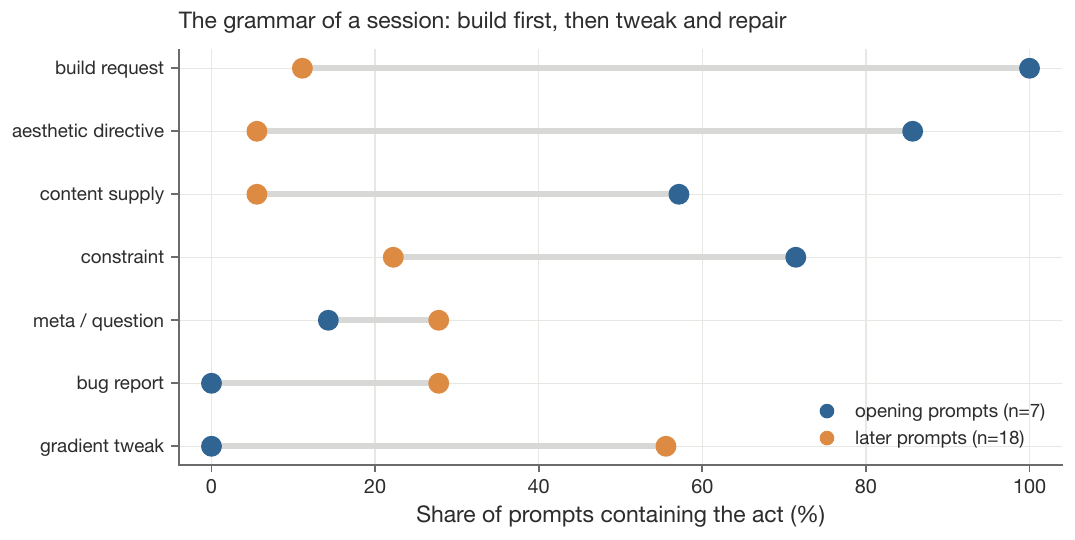}
  \caption{The grammar of a session. Speech acts in opening versus later prompts: build requests, content supply, and constraints give way to gradient tweaks and bug reports.}
  \Description{Paired bars comparing the distribution of prompt speech acts between opening and later prompts.}
  \label{fig:promptgrammar}
\end{figure}

What is typed changes as the session unfolds. Opening prompts are created based on the content provided, the build requests, the aesthetic directives, and the constraints; later prompts collapse into gradient tweaks and bug reports, with build requests falling from 100\% to 11\% of prompts and gradient modifications rising from 0\% to 56\% (Figure~\ref{fig:promptgrammar}). Reference density rises from 0.79 to 1.23 per 10 words as builder and model accumulate shared context, so later prompts increasingly point at what the model already made and specify little that is new. And the gradient adjectives requested are themselves generic (\emph{professional}, \emph{modern}, \emph{varied}, \emph{inviting}) directions that other builders are requesting of the same model at the same time. The prompt ecosystem shows the same thinness from other angles, including opening prompts that paste assignment text, prompts drafted in a second AI and pasted across (``give me the detailed plan which I can directly feed''), and a failed overhaul prompt resent verbatim without a rewrite.

\subsection{The Acceptance Test}
\label{sec:acceptance}

\begin{figure}
  \centering
  \includegraphics[width=\linewidth]{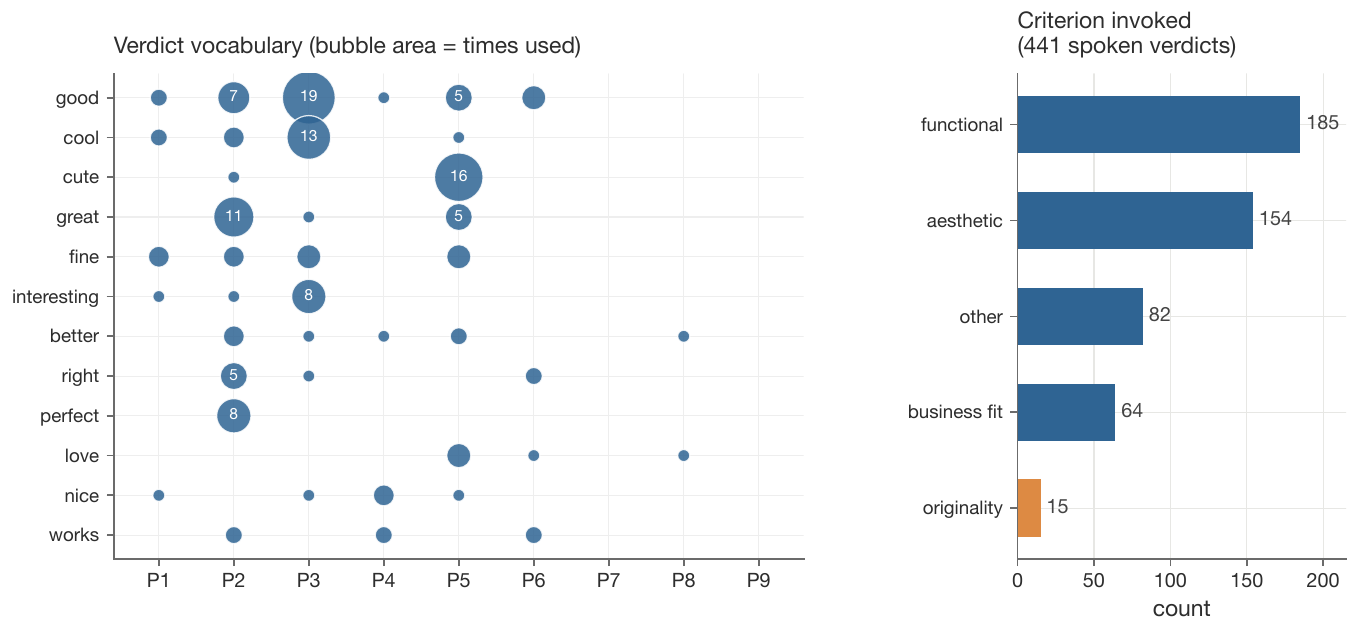}
  \caption{Verdict vocabulary and criteria. Left: top evaluative terms by builder. Right: criterion invoked across the 440 spoken verdicts; originality accounts for 3.4\%.}
  \Description{Heatmap of evaluative terms per builder alongside a bar chart of verdict criteria with originality highlighted.}
  \label{fig:verdicts}
\end{figure}

The third question is what is required for an output to be accepted because a default survives only if the acceptance test does not detect it. The spoken record of acceptance is the verdict, and nine builders spoke 440 evaluative utterances about AI output (245 positive, 147 negative, 48 mixed). The criteria they invoke are functional (185) and aesthetic (154), far ahead of business fit (64); \emph{originality is invoked 15 times, 3.4\% of all verdicts} (Figure~\ref{fig:verdicts}). The vocabulary is strikingly generic (the pooled top terms are \emph{good}, \emph{cool}, \emph{cute}, \emph{great}, and \emph{fine}) and strikingly shared, with 49\% of each person's five most frequent verdict terms sitting inside the pooled top ten on average, a common evaluative core across nine people building nine different businesses. Verdicts are also fast and frequently absent. The median gap from submission to the first spoken verdict is 67 seconds, most of it consumed by generation itself, and 14 of 54 cycles (26\%) receive no spoken verdict at all; the output simply stands. A definitional caveat is in order. A silent cycle is one with no eval-labeled utterance, and verbalization itself declines late in sessions, so late silent accepts may reflect fading think-aloud as much as fading evaluation. We read the two engagement measures together for this reason. The implication for the homogenization argument is direct. If originality is almost never part of the operative acceptance test, its absence from the artifacts requires no further explanation, whatever the grading rubric said.

\begin{figure}
  \centering
  \includegraphics[width=\linewidth]{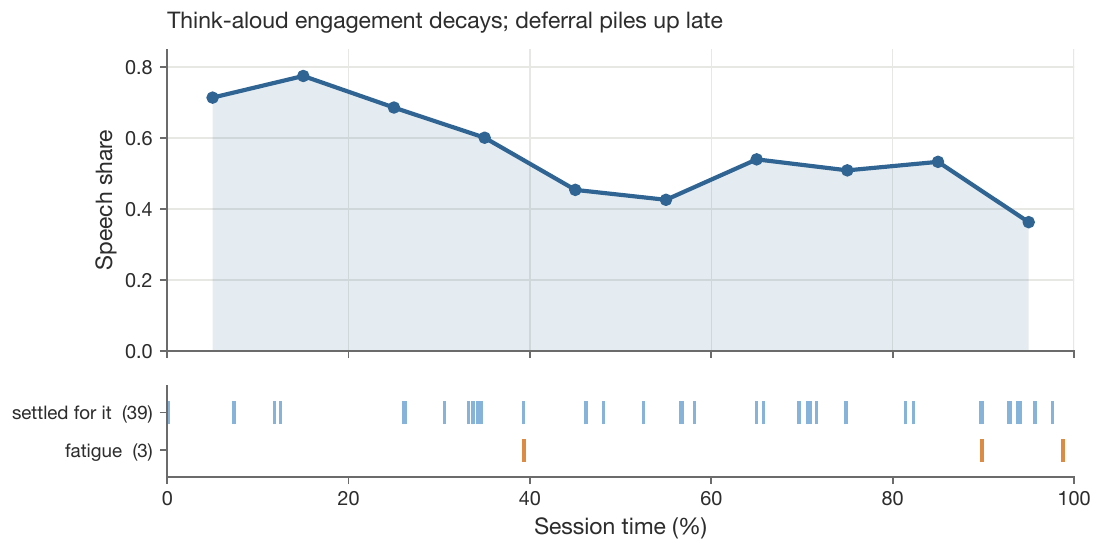}
  \caption{Fading engagement. Think-aloud speech share by session decile, with the 42 deferral moments marked on the time axis; deferrals accumulate as engagement decays.}
  \Description{Line of speech share declining across session deciles, with tick marks showing when deferral moments occur.}
  \label{fig:deferral}
\end{figure}

The test also degrades over time. There are 42 explicit deferral moments, 39 of them satisficing (``we will leave it for now''; ``that is fine'') and 3 fatigue-driven, and they pile up late in sessions, while think-aloud engagement decays in parallel, with speech covering 71\% of the first session decile and 36\% of the last. One participant explains they are out of time because they work full time; another says that if the professor tells them there is a better way, they will do it then. Acceptance late in a session therefore reflects the path of least resistance more than a considered judgment (Figure~\ref{fig:deferral}).

\begin{figure}
  \centering
  \includegraphics[width=\linewidth]{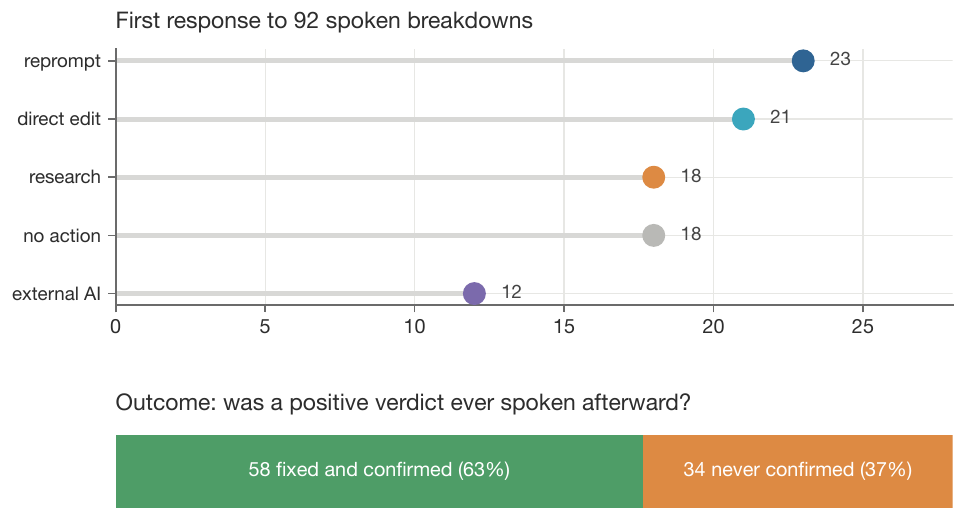}
  \caption{Breakdowns and repair. First response to the 92 spoken breakdowns and their outcomes; 37\% are never confirmed fixed.}
  \Description{Flow from spoken breakdowns through first repair actions to confirmed-fixed versus never-confirmed outcomes.}
  \label{fig:repair}
\end{figure}

Even visible failures pass. Builders verbally flagged 92 breakdowns; the first response splits almost evenly across reprompting (23), direct editing (21), researching (18), no action at all (18), and going to an external AI (12). The outcome distribution is the finding, as 34 of 92 (37\%) are never followed by a positive verdict; the problem is shipped or silently dropped (Figure~\ref{fig:repair}). Defects survive by the same route as defaults, through acceptance without verification.

Together, the three subsections document the conditional chain mentioned in the introduction. Since the loop is one script, the same prior is summoned in the same way for everyone. This is why convergence takes the form of recurring archetypes. Human input into the script is minimal and decreasing, so the unspecified space that the defaults determine is significant. The acceptance test is generic, fast, and often absent. It is almost never about originality, so what the defaults decide stays unexamined. Each link is documented as present; Section~\ref{sec:whatwelearned} states precisely what this does and does not establish. One element of the mechanism remains, namely why none of this feels like homogenization from the inside.

\section{Results III: The Convergence Is Invisible to Its Creators}
\label{sec:invisible}

The remaining question is RQ3: whether anyone could \emph{tell}, including the escapees. The process layer shows the invisibility being produced in real time; the perception layer confirms it at cohort scale.

\subsection{Invisibility in Production: Narrating the Doing, Crediting the Making}
\label{sec:agency}

\begin{figure}
  \centering
  \includegraphics[width=\linewidth]{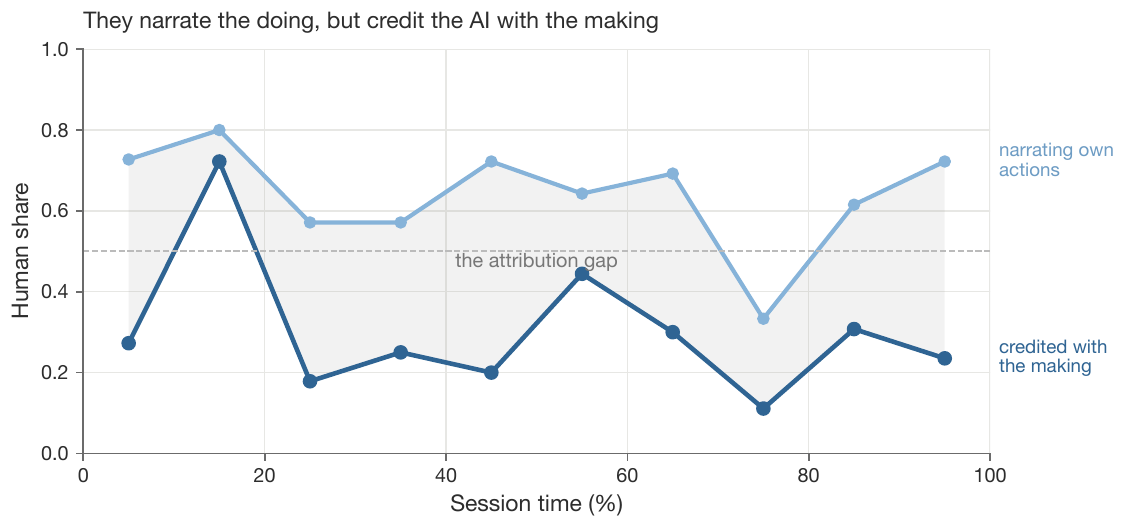}
  \caption{Narrating the doing, crediting the making. Human share of creation language across session time: narration of action (light) versus attributed credit for the making (dark). Builders narrate in the first person while crediting the artifact to the AI.}
  \Description{Two lines over normalized session time, showing the human share of action narration stays high while the human share of attributed credit stays low and declines.}
  \label{fig:agency}
\end{figure}

The two agency measures point in opposite directions, and the gap between them is the finding. In raw narration, builders talk in the first person, with 131 first-person constructions with creation verbs against 69 with the AI as subject (human share 0.66). In attributed credit (who is said to have \emph{made} the thing), the AI dominates at 130 against 66 (human share 0.34), and the human share drifts down across the session, from 0.38 to 0.28 (Figure~\ref{fig:agency}). Builders experience the session as their own activity while attributing the artifact to the model: the experience of agency is manufactured by the doing (typing, waiting, peeking, judging) \citep[cf.][]{pierce2001}, while the making, where the design decisions happen, is accurately felt to sit with the AI. A builder in that position has every reason to feel authorial and no occasion to audit the design decisions they never made. This is the on-camera, utterance-level signature of the decoupling the survey now confirms at scale.

\subsection{Invisibility at Scale: Felt Authorship, Satisfaction, and Quality Do Not Track Originality}
\label{sec:decoupling}

Builders felt in charge, with mean felt authorship at 67/100 (``I led, not the AI''). They were also pleased, with mean satisfaction with the AI at \rr{6.4}/7 and mean perceived quality of their own site at 6.1/7. These judgments were internally coherent, with satisfaction, perceived quality, and information foraging intercorrelating at $r = \rr{0.47}$--$\rr{0.56}$ (all $p < .001$).

\begin{figure*}
  \centering
  \includegraphics[width=\linewidth]{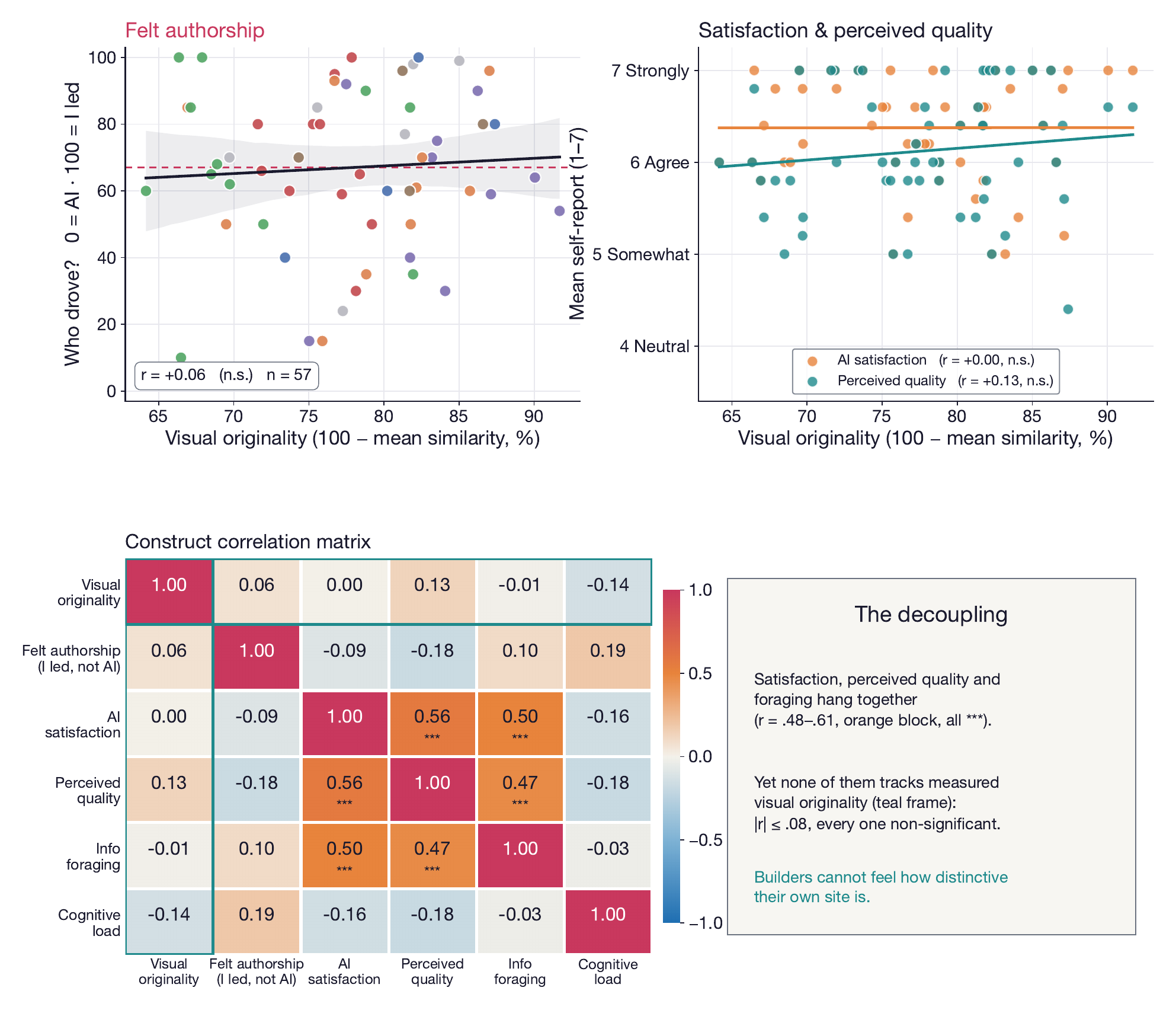}
  \caption{The perception--originality decoupling. Top: builders' self-reports
  plotted against the measured visual originality of their sites (100 minus mean
  similarity to the rest of the cohort). Felt authorship (left; ``who drove the
  build,'' 0--100) and AI satisfaction and perceived quality (right; 1--7) are
  all flat against originality --- $r = +0.06$, $+0.00$, and $+0.13$
  respectively, none significant ($n = 57$--$58$); points in the left panel are
  colored by archetype, and lines are OLS fits with bootstrap 95\% confidence
  bands. Bottom: the construct correlation matrix. The three subjective measures
  intercorrelate strongly ($r = 0.47$--$0.56$, all $p < .001$; orange block),
  yet none correlates with visual originality (teal frame). Builders' felt
  experience of vibe coding thus carries no information about the objective
  distinctiveness of what they produced. Stars denote $^{*}p<.05$, $^{**}p<.01$,
  $^{***}p<.001$.}
  \Description{A felt-authorship scatter and a satisfaction/quality scatter, both
  flat against measured originality, above a six-by-six construct correlation
  matrix in which the three subjective measures correlate strongly with one
  another but not with visual originality.}
  \label{fig:decoupling}
\end{figure*}

None of them tracks measured originality. The correlation with felt authorship is $r = \rr{+0.06}$, 95\% CI $\rr{[-0.20, +0.32]}$ ($n = \rr{57}$); with satisfaction, $r = \rr{+0.00}$, 95\% CI $\rr{[-0.26, +0.26]}$; with perceived quality, $r = \rr{+0.13}$, 95\% CI $\rr{[-0.14, +0.37]}$ ($n = \rr{58}$ for the latter two items; Figure~\ref{fig:decoupling}).\footnote{Confidence intervals and equivalence margins are computed from the reported correlations and sample sizes via the Fisher transform.}

Equivalence testing bounds the null without establishing it. TOST supports equivalence to zero only for margins of $|r| \geq 0.27$ (felt authorship), $0.24$ (satisfaction), and $0.29$ (perceived quality); these are the minimal margins derived from the 90\% confidence intervals, and no smallest effect size of interest was prespecified. The data are therefore inconsistent with moderate or larger relationships but cannot rule out small ones. Under the range restriction documented below, even these margins are optimistic, since the disattenuated equivalence bounds are wider. The builders who escaped the house styles felt no more authorial, no more satisfied, and no prouder than the builders sitting in the center of the largest cluster. Recall that every one of them was instructed, and graded, to avoid generic AI aesthetics: the incentive to be original was explicit, the belief in having been original was strong, and the relationship between that belief and the fact was indistinguishable from zero within these bounds.

This bounded null is the finding, and we frame it with its alternatives in view. The subjective measures are internally coherent and intercorrelated, and the originality measure is conservative (Section~\ref{app:robustness}). Two psychometric alternatives nonetheless require explicit treatment before the structural interpretation can stand. Felt authorship is a single 0--100 item, and single-item unreliability attenuates correlations; the originality measure is also range-restricted in this cohort (the spectrum in Figure~\ref{fig:ranking} runs roughly 64--96 with most of its mass between 70 and 85), which attenuates further. What survives under any of these readings is the absence of a detectable signal exactly where the incentive structure says one should be; what the structural account adds is a reason to expect that absence, beyond merely observing it. The data show a \emph{decoupling}, in which the subjective experience of vibe coding (agency, satisfaction, pride) carries no detectable information about the objective distinctiveness of its output.

\subsection{The Intention Was Present: Evidence from the Written Reflections}
\label{sec:reflections}

A fourth data source shows that the missing correlation from the previous section cannot be explained by indifference. After completing the site, each student wrote a one- to two-page reflective blog post in a natural style, as if sharing the experience on a professional network. The corpus comprises 78 usable posts (median length: 821 words) from 83 students who submitted work, 72 of whom were builders in the artifact cohort. Since the posts were graded coursework intended for the instructor, we interpreted them as statements of intention and self-presentation rather than unguarded testimony.

The reflections show that avoiding genericity was a live, articulate concern. Forty-eight of the 78 posts discuss originality, distinctiveness, or the generic look of AI output, and many describe deliberate tactics. A student building a barbershop site ``intentionally avoided flashy colors and generic AI visuals.'' Another, building for an architecture studio, chose ``whitespace, a consistent color palette, and simple typography to avoid a generic `AI-built' look.'' A student working on a dance studio wrote that ``sounding generic is basically the same as being invisible.'' A photographer's site was judged finished when it ``stopped looking like a generic creative portfolio and started looking like it belonged to a photographer.''

The reflections also rehearse the attribution pattern of Section~\ref{sec:agency} in a considered register. Forty-seven posts discuss who did the making. One student describes ``shifting my role from technical builder to creative director''; another writes that the model ``handled the heavy lifting of responsive layouts and initial styling'' while ``100\% of the site's `soul' came from human direction.'' Forty-three posts volunteer a numerical split, most commonly attributing 60 to 70 percent of the initial output to the model and the remainder to their own direction. The same builders who credit the model with the fabric of the artifact claim its identity for themselves, which is felt authorship in the exact sense the survey measures.

Read against Results I, the reflections rule out one explanation for the missing correlation of Section~\ref{sec:decoupling}: indifference. If builders had not cared about genericity, or had not believed they avoided it, the absence of a correlation would follow trivially. The reflections show the opposite. The intention to be distinctive was specific and widely shared, and the belief that it had been achieved was strong. Yet neither the intention nor the belief predicts a site's measured originality. Builders were not short on motivation. What they lacked was the comparison set that originality is defined against, the distribution of every other builder's site, which only the platform can observe. The next section names this condition: authored ignorance.

\subsection{Authored Ignorance}
\label{sec:aai}

The mechanism, we argue, is structural. Its core is positional, since originality is a relational property defined against the distribution of what everyone else produced. No individual builder observes that distribution. Each sees one artifact, their own, looking professional, responsive, and aligned with their intent. The judgment ``this avoids generic AI aesthetics'' requires exactly the comparison set that is unavailable to the person making it, and fully available to the platform. This is the structural core of what we call \emph{authored ignorance}: the builder is ignorant of where the artifact sits, and that ignorance follows from the viewing position itself. It admits a specific remedy. Position cannot be repaired by inspecting one's own site; repair requires revealing the distribution to which the site belongs. The process data exposes a second, finer-grained layer. Even within the single artifact a builder can see, the design decisions were made by the model, narrated by the human, and credited away in real time (Section~\ref{sec:agency}), so no moment in the loop makes auditing those decisions anyone's job. We call the full phenomenon \emph{authored ignorance}; in the visual-design setting studied here, its object is aesthetic, namely the felt authorship of a design whose genericity its author is structurally unable to perceive.

\section{Discussion}

\subsection{What We Learned}
\label{sec:whatwelearned}

Seventy-three builders with seventy-three different design problems used one tool and produced roughly a dozen distinct designs, while reporting high authorship and satisfaction, and despite explicit instructions to avoid generic AI aesthetics. The three results sections establish, in turn, the convergence, its candidate mechanism, and its invisibility. The convergence itself (Results~I) shows that the homogenization documented for AI-assisted text extends, with at least equal force, to multimodal design artifacts produced by full delegation. The mechanism (Results~II) documents the candidate chain of one script summoning the same defaults identically for everyone, specification thin enough to leave most of the design space to the stack, and an acceptance test that almost never asks the one question the assignment graded. The invisibility (Results~III) shows why market feedback is unlikely to correct it. The people best positioned to demand diversity, the builders themselves, cannot perceive its absence, and the process data shows that incapacity being produced, as credit for the making is handed to the model in real time, in sessions whose pace leaves no moment for auditing the design decisions.

A scope note on the word \emph{mechanism} is owed before the comparison to prior work. The process layer documents every link of the chain as \emph{present} (the shared loop, the thin specification, the shallow test, the misattributed credit), and it documents them in the same cohort whose outputs converged. It does not show any link causing the next, since with nine builders we deliberately abandoned between-person hypothesis tests; our design observes the candidate mechanism in operation and does not test whether it produces the outcome. Coherence and completeness license the mechanism language, while identification awaits an experimental design. Every condition the default-capture account requires is observed, none is contradicted, and the account predicted the perception results of Results~III in advance. A design that manipulates a link (imposed specification, forced comparison at acceptance) is the causal test, and Section~\ref{sec:limits} names it.

The candidate mechanism differs from prior text-domain findings in an instructive way. In story-writing experiments, homogenization travels through \emph{anchoring}: the human author absorbs the model's suggestion and writes toward it. In vibe coding there is no separate human composition for the suggestion to anchor; the production stack's defaults \emph{are} the artifact unless contested, and the process data measures how rarely the contest happens, with voiced intentions dropped between mind and keyboard, originality invoked in 3.4\% of verdicts, and a quarter of generations accepted in silence. This explains both the magnitude of convergence in a setting with maximal input diversity, and its structure, which favors recurring archetypes over diffuse similarity, because the defaults are themselves discrete styles the stack returns to, and the one script queries them the same way every time.

\subsection{Theoretical Contribution}\label{sec:fluency}

For the creativity-and-AI literature, the results extend the individual/collective tension \citep{doshi2024, boussioux2024crowdless} to a domain where the collective cost may bind hardest, since visual identity is a positional good, valuable precisely insofar as it differs from neighbors. The results also connect this literature to the older homogenization-of-the-web finding \citep{goree2021}: the convergence that shared frameworks produced gradually over a decade, a shared generative stack reproduces within a single cohort, with the human composition step that previously preserved variation no longer present. In the stage-dependency framing of \citet{boussioux2026hidden}, vibe coding is the worst case, because the AI occupies every stage from ideation through execution and leaves no human-led phase to seed diversity. In interaction-design terms, the regime widens both of Norman's gulfs while subjectively narrowing them \citep{norman1986}. Execution feels effortless because it is only typing, and evaluation feels conclusive because the artifact arrives finished. Yet what is lost on each side, the untyped intention (Section~\ref{sec:thininput}) and the unasked originality question (Section~\ref{sec:acceptance}), is exactly where the defaults enter and where they persist.

For theories of human-AI authorship, the decoupling result generalizes the ghostwriter effect \citep{draxler2024} in an instructive direction. That dissociation concerns ownership of an artifact the author can fully see; authored ignorance, as we define it, concerns a property the author \emph{cannot} see at all, the artifact's position in a distribution only the platform observes. The process layer supplies the microfoundation, as the attribution of making drains to the model in real time even as the narration of doing stays first-person (Section~\ref{sec:agency}), so the ignorance is a live property of the loop, visible before any survey is administered. The root is single. Vibe coding manufactures the experience of creation while withholding the information that creation, in the evaluative sense, requires.

The account also explains why nobody in the loop objects. Typical
designs are fluent to process, and that fluency reads as liking:
prototypicality, measured as proximity to a category average, predicts
car sales \citep{landwehr2011}, and the preference for typicality
reverses only at higher exposure, where atypical designs gain
\citep{landwehr2013}. Our acceptance data sits in the low-exposure
regime. Builders reach a first verdict a median of 67 seconds after
submitting, inspect in nine-second glances, and accept a quarter of
generations without a word (Section~\ref{sec:acceptance}). A verdict
formed under those conditions is a fluency judgment, and fluency is what
the stack's defaults maximize. Genericity is not a failure of the loop
but the thing the loop selects for.

The mismatch is temporal. The builder judges the site once. The business
lives with it at repeated exposure, where differentiation is what pays.
Nobody in the loop holds that longer position. Not the builder, who
accepts on sight. Not the model, whose training rewards output people
approve of immediately \citep{xiao2025collapse}. And not the
stakeholders: across the 65 documented feedback sessions of
Appendix~\ref{app:stakeholders}, owners asked for accurate hours and
working links and customers asked for usability repairs, while no one
asked the site to look less like other sites. They were first-time
viewers too. The artifact passes through a chain of single-exposure
judges, and the property that only repeated exposure rewards has no
advocate in it.

\subsection{Implications for Platform Design}

The diagnosis implies the remedy, because the missing information already exists and is cheap to surface. Table~\ref{tab:implications} maps each documented mechanism to its design response; we develop each below.

\begin{table}
  \caption{From documented mechanism to design response.}
  \label{tab:implications}
  \small
  \begin{tabular}{@{}p{0.46\linewidth}p{0.46\linewidth}@{}}
    \toprule
    Documented mechanism (evidence) & Design response \\
    \midrule
    Originality is relational; the distribution is visible only to the platform (Sections~\ref{sec:decoupling}--\ref{sec:aai}) & \textbf{Originality feedback}: surface the artifact's distributional position at generation time \\
    One shared generate-and-check script; a single candidate per request (Section~\ref{sec:onescript}) & \textbf{Divergence-promoting generation}: sample initial candidates from distinct archetypes \citep{dow2010} \\
    Thin and thinning specification; fast, generic, often absent acceptance; 61\% of generation time is dead waiting (Sections~\ref{sec:onescript}--\ref{sec:acceptance}) & \textbf{Deliberate friction}: pre-generation impositions and comparison work, funded by the dead wait \\
    The effective number of distinct designs is measurable from the platform's own outputs (Section~\ref{sec:archetypes}) & \textbf{Diversity as a platform metric}: track the effective number over time \\
    \bottomrule
  \end{tabular}
\end{table}

\textbf{Deliberate friction.} This is the best-grounded implication because the process data identifies where friction would occur and the loop supplies the budget to pay for it. A substantial share of voiced intentions never reaches the prompt (Section~\ref{sec:thininput}), and the acceptance test rarely examines what arrives (Section~\ref{sec:acceptance}). Small impositions before generation, such as requiring a visual reference, a palette choice, or a ``what should this \emph{not} look like'' prompt, target exactly the unspecified space the defaults currently claim; and 61\% of generation time is dead waiting (Section~\ref{sec:onescript}), attention the platform is already consuming and could repurpose into specification or comparison work at no net cost to speed.

\textbf{Originality feedback.} The platform computes, or could compute, exactly the statistic our pipeline does. An originality meter (``layouts like this one were generated 4,000 times this week'') converts the invisible distribution into a design signal, the way plagiarism detectors did for text. Which distribution the meter should display is itself a design question. A restaurant competes visually with the restaurants its customers also see, and locally or sectorally conditioned comparison sets may therefore serve differentiation better than global ones.

\textbf{Divergence-promoting generation.} Platforms typically return one design. Sampling initial candidates from \emph{distinct archetypes}, which is feasible since the archetypes are recoverable from the stack's own outputs, would relocate the human from approver of a default to selector among directions, restoring a human-led ideation stage; presenting parallel alternatives is among the best-evidenced interventions in design research, increasing divergence and self-efficacy without harming outcomes \citep{dow2010}. The implication is conditional on where the archetypes live: if they prove to reside mainly in the platform's scaffolding and less in the model's learned priors (Section~\ref{sec:limits}), the same goal is served by diversifying the template layer instead.

\textbf{Diversity as a platform metric.} At the ecosystem level, the effective number of distinct designs is a measurable, optimizable quantity. A platform that tracked it would know, before the web does, whether it is becoming a monoculture. The stakes compound over time, because generated artifacts re-enter the corpora on which future models train, and training on model-generated data measurably degrades performance and lexical diversity \citep{zhang2026}.

\subsection{Limitations and Future Work}
\label{sec:limits}

Our cohort is one class, one recommended platform, one model generation. The specific archetypes will drift as models update, though the structural argument (shared defaults plus invisible distributions) does not depend on which styles recur. Graduate students are not professional designers, but they are arguably \emph{the} vibe-coding population, namely capable non-specialists for whom the tools were built. Originality here is visual; a site can be visually generic and verbally distinctive. The survey joins 58 of 73 sites ($n = 57$ for the authorship item), and our null results bound the correlations without establishing equivalence (Section~\ref{sec:decoupling}), with single-item measurement and range restriction as live attenuation channels. The process layer is nine builders: all process claims are made at the cycle, episode, or utterance level with sessions as replications, and we draw no between-person inferences from it. Its utterance- and prompt-level codes are AI-assisted first passes whose stratified human validation is pending (Section~\ref{sec:process}). Generation ends are behaviorally inferred, so second-level latencies around them carry construction noise. Think-aloud does not exhaust intention, so the translation-loss rate conditions on verbalized intent. Appendix~\ref{app:robustness} localizes the convergence within the page: hero regions on their own are markedly less convergent than whole pages, so much of the shared structure sits below the fold. Our explanans is the production stack as a whole. The archetypes may live in the model's priors, the platform's scaffolding, or their interaction, and narrow priors have independent support, since alignment training can concentrate a model's output distribution to the point of preference collapse \citep{xiao2025collapse}. Controlled probes (prompting the raw model without the platform, or holding the platform fixed across model versions) would disentangle them, and the design implications split accordingly. The translation-loss rate also lacks a cross-medium anchor; a comparison condition in which builders brief a human developer or complete a structured intake would calibrate it. Finally, our identification is design-based. The cohort design removes brief and builder heterogeneity but shares a genre, an assignment scope, and a course context, which the comparison corpus (Section~\ref{sec:baseline}) anchors. A randomized comparison across platforms, and a manipulation of a mechanism link such as imposed specification or forced comparison at acceptance, are the natural next studies.

\section{Conclusion}

The long tail of the web is expected to reflect the unique, localized, and diverse nature of the businesses it represents. In our cohort, vibe coding made building accessible to everyone, but it left most builders with similar sites. The similarity was invisible, above all, to the builders themselves. The question the baseline and replication agenda above is built to answer is whether the web at large follows. The structural argument of shared defaults and invisible distributions suggests that it will follow this pattern wherever the production stack is shared. We do not interpret these results as a case against AI. The solution lies in platforms that show builders where their designs stand in a distribution that only the platform can see and in interaction loops that prompt builders to provide specifications before acceptance.



\bibliographystyle{ACM-Reference-Format}
\bibliography{references}

\appendix

\section{Measurement Robustness Details}
\label{app:robustness}

This appendix collects the full numbers behind the robustness statements of Sections~\ref{sec:method} and~\ref{sec:archetypes}. All computations use the primary DINOv3 (ViT-B/16) native-resolution embeddings of the 73 homepage screenshots; scripts and outputs accompany the archival package.

\subsection{Clustering Sensitivity}

\begin{figure}[t]
  \centering
  \includegraphics[width=\linewidth]{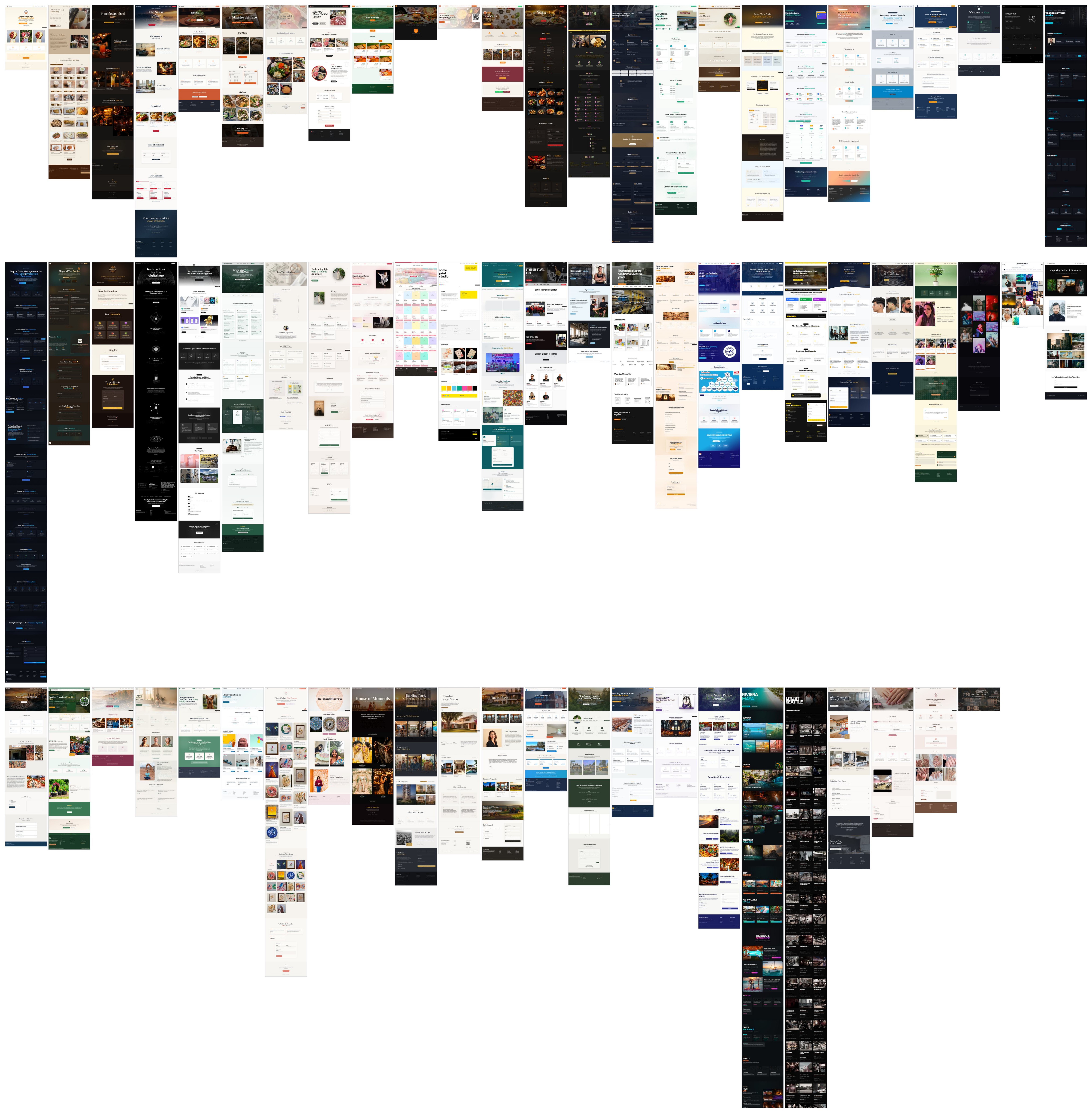}
  \caption{\rr{The design space as a similarity wall. All 73 full-page screenshots, rendered as full-height strips and placed in a single one-dimensional order (spectral seriation of the DINOv3 cosine-similarity matrix, refined by a 2-opt pass) so that visually similar pages are adjacent; read left-to-right, top-to-bottom. Neighboring pages share layout, palette, and section rhythm, and the dense image-heavy portfolios cluster tightly at one end.}}
  \label{fig:fullpagewall}
\end{figure}

Density clustering on 73 points is configuration-sensitive, so we treat the archetype partition as descriptive and report its stability explicitly. A grid over 60 UMAP--HDBSCAN configurations (UMAP neighbors $\in \{10, 12, 15, 20, 25\}$, minimum distance $\in \{0.05, 0.10, 0.12, 0.25\}$, HDBSCAN minimum cluster size $\in \{3, 4, 5\}$; fixed seed) yields archetype counts from 3 to 12 (median 7), with 78--100\% of sites assigned. We exclude minimum cluster size 2 on the grounds that two-site clusters are pairs, not archetypes. The production configuration (neighbors 20, minimum distance 0.12, minimum cluster size 4, seed 42) yields the six archetypes of Figure~\ref{fig:housestyles}, absorbing 64 of 73 sites (88\%). Exact partitions are additionally sensitive to library internals and can vary across library versions even at a fixed random seed; the pipeline therefore ships with pinned dependency versions, the partition is used descriptively, and the quantitative claims rest on the clustering-free statistics below. The full similarity structure that these partitions summarize is shown as a seriated wall of all 73 pages in Figure~\ref{fig:fullpagewall}.

\subsection{Effective Number of Distinct Designs}

The order-$q$ Vendi score \citep{friedman2023} is the exponential of the R\'enyi entropy of the eigenvalue spectrum of the normalized similarity kernel $K/n$; it generalizes ecology's effective-number-of-species indices to a similarity kernel and requires no clustering, projection, or tunable parameters. On the cosine-similarity kernel of the 73 sites, the order-2 score is 12.0. Its leave-one-out range is $[11.7, 12.4]$; a jackknife 95\% confidence interval is $[9.1, 15.0]$, and subsampling (66 of 73 sites without replacement, 1{,}000 replicates) gives $[11.0, 12.8]$. We report both interval constructions because neither is canonical for spectral statistics; the na\"ive bootstrap-with-replacement is biased downward for diversity indices (duplicated rows collapse eigenvalue mass) and is not used. Occupancy-based counterparts on the six-archetype partition, treating each unclustered site as a singleton, give 6.0 (inverse-Simpson) and 8.0 (Shannon). Under the fixed-depth crops below, the order-2 Vendi score ranges from 12.0 (full page) to 16.3 (shallowest crop).

\subsection{Fixed-Depth Crops and Page Length}

Full pages range from 720 to 17,581 pixels tall (median 5,010) at a fixed 1{,}280-pixel width. Per-site originality correlates with page height ($r = -.35$, $p = .002$), so we re-embedded every page cropped to a common maximum depth through the identical pipeline (pages shorter than the cap keep their full height: 3, 11, and 30 of 73 pages at the three depths).
Table~\ref{tab:fixdepth} reports the result. Deeper crops are steadily more similar, which places much of the shared structure below the fold, and the per-site ordering is only moderately preserved at the shallowest depth ($r = .62$).

\begin{table}[h]
\caption{\rev{Fixed-depth crop robustness. Convergence statistics when every page is embedded at a common maximum scroll depth.}}
\label{tab:fixdepth}
\rev{\begin{tabular}{lccccc}
\toprule
Depth & Mean sim. & Median NN & Pairs $>$90\% & Originality corr.\ w/ full page & Orig.\ vs.\ height \\
\midrule
1,440\,px & \rr{17.2\%} & \rr{38\%} & 0 & $r=\rr{.62}$ & $r=-.09$ (n.s.) \\
2,880\,px & \rr{20.9\%} & 45\% & 0 & $r=\rr{.87}$ & $r=\rr{-.23}$ \\
4,320\,px & \rr{22.2\%} & \rr{48\%} & 0 & $r=.98$ & $r=\rr{-.28}$ \\
full page & \rr{22.8\%} & \rr{49\%} & 0 & --- & $r=\rr{-.35}$ \\
\bottomrule
\end{tabular}}
\end{table}

The headline statistics are essentially unchanged at depths that cover most pages, and the length association weakens toward zero under the strictest geometric equalization. The residual association at intermediate depths is consistent with the substantive reading that default-accepting builders receive the stack's long, many-sectioned template, but our design cannot separate that reading from residual geometric sensitivity, and we treat it as a hypothesis.

\subsection{Hero-Region Variant}

Embedding only the above-the-fold hero region (top $1280 \times 720$ crop, identical native-resolution preprocessing) yields mean pairwise similarity \rr{11.2\%}, median nearest-neighbor similarity \rr{28\%}, and an order-2 Vendi score of \rr{24.2}, against \rr{22.8\%}, \rr{49\%}, and \rr{12.0} for full pages. Because absolute cosine levels shift with input content statistics across crop types, the comparison rests on its consistent direction across all three statistics, not on any single number. 

\subsection{\texorpdfstring{Hero Archetypes: Layout $\times$ Palette}{Hero Archetypes: Layout x Palette}}
\begin{figure*}[t]
  \centering
  \includegraphics[width=\linewidth]{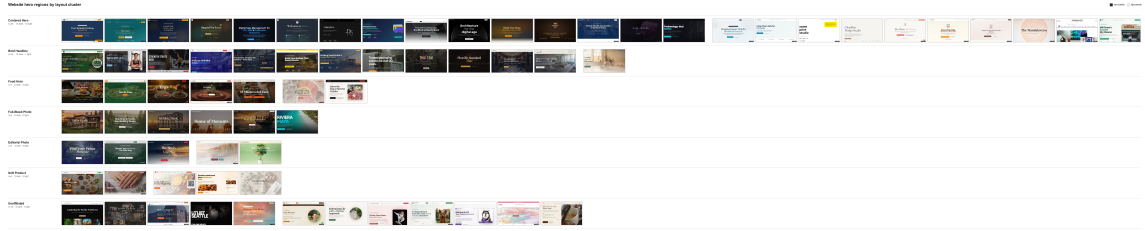}
  \caption{Hero archetypes $\times$ palette. All 73 hero regions grouped by their six hero-layout clusters (plus the unaffiliated group); within each row the dark-palette members are on the left and the light-palette members on the right, and members are ordered so visually similar heroes sit adjacent. The same layout skeleton recurs across both dark and light skins, and some layouts (e.g.\ the full-bleed photo hero) are almost entirely dark---palette is a skin applied across layouts rather than an archetype of its own.}
  \label{fig:heromontage}
\end{figure*}

\rr{Clustered on the hero-region embeddings with the same UMAP--HDBSCAN grid used for the full-page partition, but with the configuration the hero count-selection diagnostics (DBCV and grid-consensus stability) actually favor---neighbors 25, minimum distance 0.12, minimum cluster size 5, seed 42---the 73 heroes fall into six recurring layouts (61 of 73 assigned, 84\%; the full-page production setting is min-cluster-size 4, which for the hero embeddings yields a coarser five-cluster split). Because the DINOv3 embedding is dominated by layout and composition and is largely invariant to global palette, a single hero layout recurs in both a dark and a light skin: splitting each site by mean hero luminance (dark below 160, light at or above, on a 0--255 scale; the luminance distribution is bimodal with a sparse 120--180 gap) gives the layout $\times$ palette structure of Table~\ref{tab:heropalette} and Figure~\ref{fig:heromontage}. Across the cohort, 48 heroes (66\%) are dark and 25 (34\%) light, and the largest layout---a centered-tagline hero with navigation and a call-to-action---appears as 15 dark and 10 light instances. The convergence therefore operates at two levels: the production stack reuses a small set of layout skeletons and reskins them across a narrow palette, so palette variety does not imply layout variety.}

\begin{table}[h]
\caption{Hero-region layout archetypes (rows, by size) $\times$ palette (dark/light by hero luminance), 73 sites.}
\label{tab:heropalette}
{\begin{tabular}{lccc}
\toprule
Hero layout archetype & Dark & Light & Total \\
\midrule
Centered Hero (headline + nav + CTA) & 15 & 10 & 25 \\
Bold Headline (text-forward) & 12 & 1 & 13 \\
Food Hero (food photography) & 5 & 2 & 7 \\
Full-Bleed Photo (edge-to-edge image) & 6 & 0 & 6 \\
Soft Product (light, product/lifestyle) & 2 & 3 & 5 \\
Editorial Photo (atmospheric + serif) & 3 & 2 & 5 \\
Unaffiliated & 5 & 7 & 12 \\
\midrule
Cohort & 48 & 25 & 73 \\
\bottomrule
\end{tabular}}
\end{table}

\begin{figure*}[t]
  \centering
  \includegraphics[width=\linewidth]{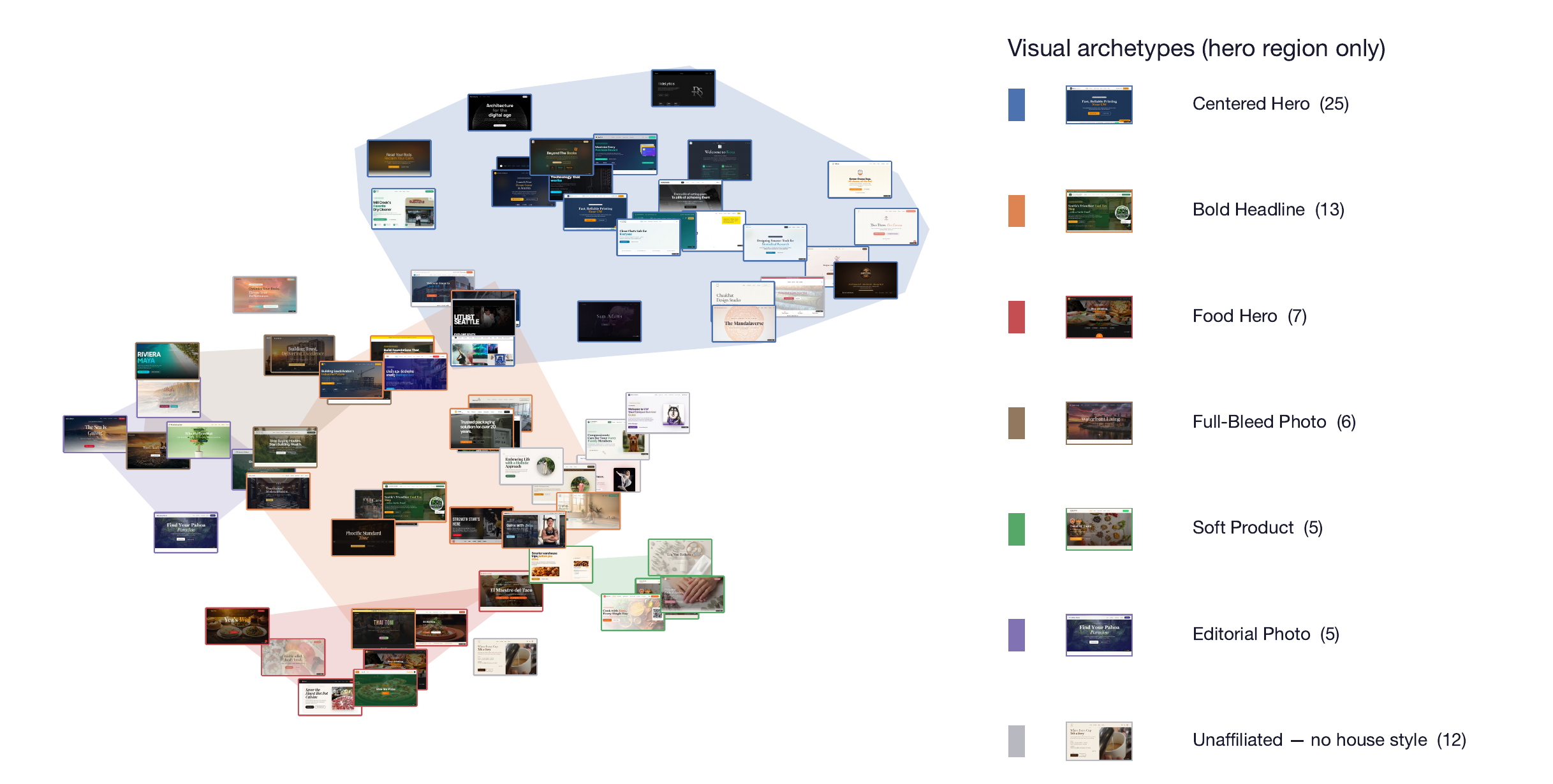}
  \caption{\rr{The hero-region design space. UMAP projection of the hero-only DINOv3 embeddings, each site drawn as its hero thumbnail and framed by its hero-layout archetype (the six of Table~\ref{tab:heropalette}, config neighbors 25 / min\_dist 0.12 / min-cluster-size 5). Compare Figure~\ref{fig:map}, which maps the full-page embeddings.}}
  \label{fig:maphero}
\end{figure*}

\begin{figure*}[t]
  \centering
  \includegraphics[width=\linewidth]{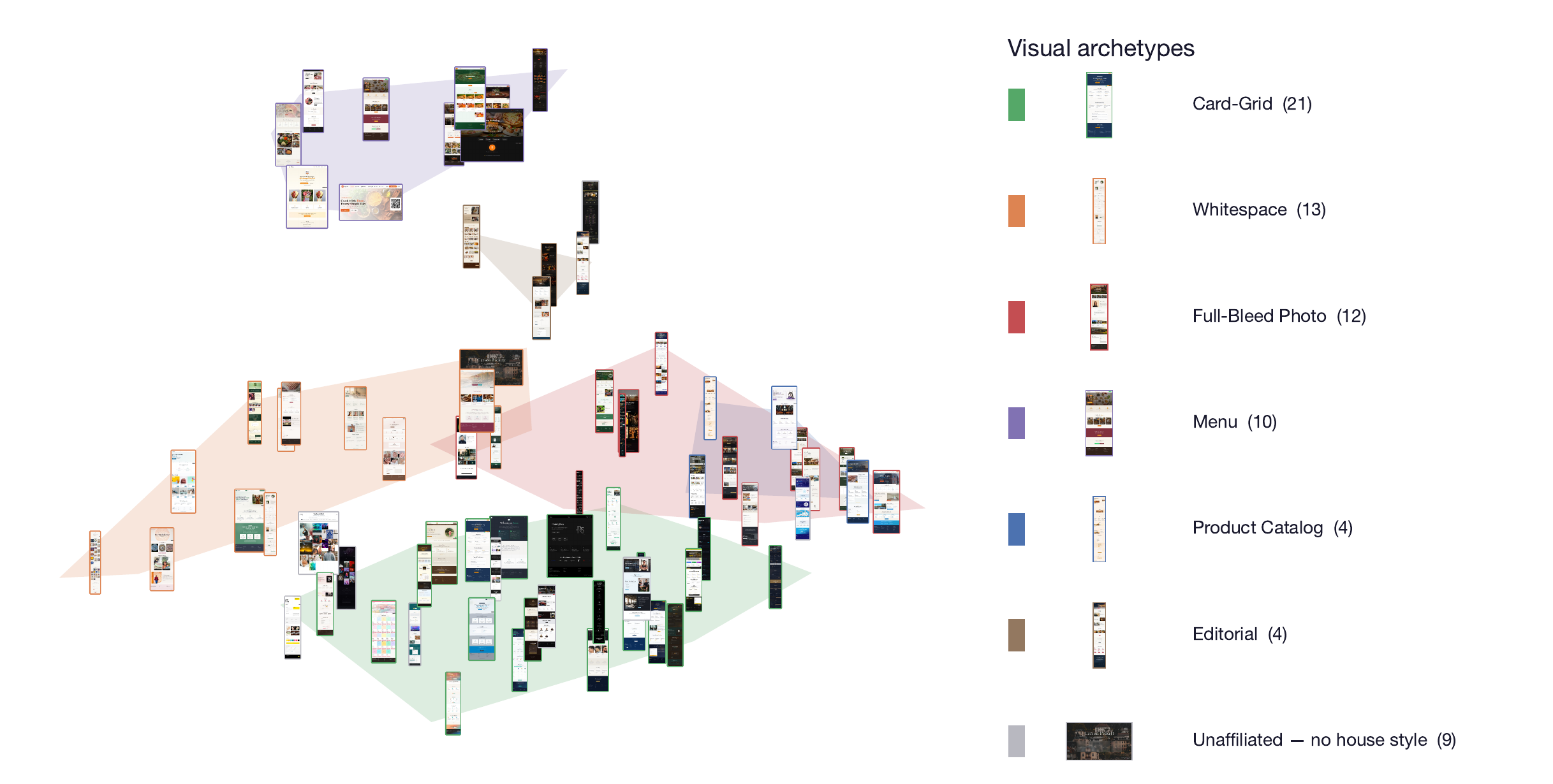}
  \caption{\rr{The full-page design space, drawn with full-page thumbnails. The same UMAP projection and six full-page archetypes as Figure~\ref{fig:map} (Card-Grid, Whitespace, Full-Bleed Photo, Menu, Product Catalog, Editorial), but each site is shown as its entire page rather than a hero crop, so the recurring section rhythms below the fold are visible.}}
  \label{fig:mapfullpage}
\end{figure*}

\subsection{\texorpdfstring{\rr{Color Palettes and Their Convergence}}{Color Palettes and Their Convergence}}

\rr{To examine palette as its own axis, we extract each site's color palette by $k$-means ($k=5$) on the full-page screenshot, yielding five dominant colors with their screen-area weights; a color is \emph{neutral} if its HSV saturation is below 0.15 (white, black, gray, cream), and a site's \emph{brand accent} is the highest-area vivid mid-tone (saturation $\ge 0.30$, lightness between 15 and 90 on a 0--100 scale). Because extraction is on the rendered page, photographic imagery contributes to the palette; this is faithful to what a viewer sees but blends brand chrome with photo content.}

\begin{figure}[h]
  \centering
  \includegraphics[width=0.72\linewidth]{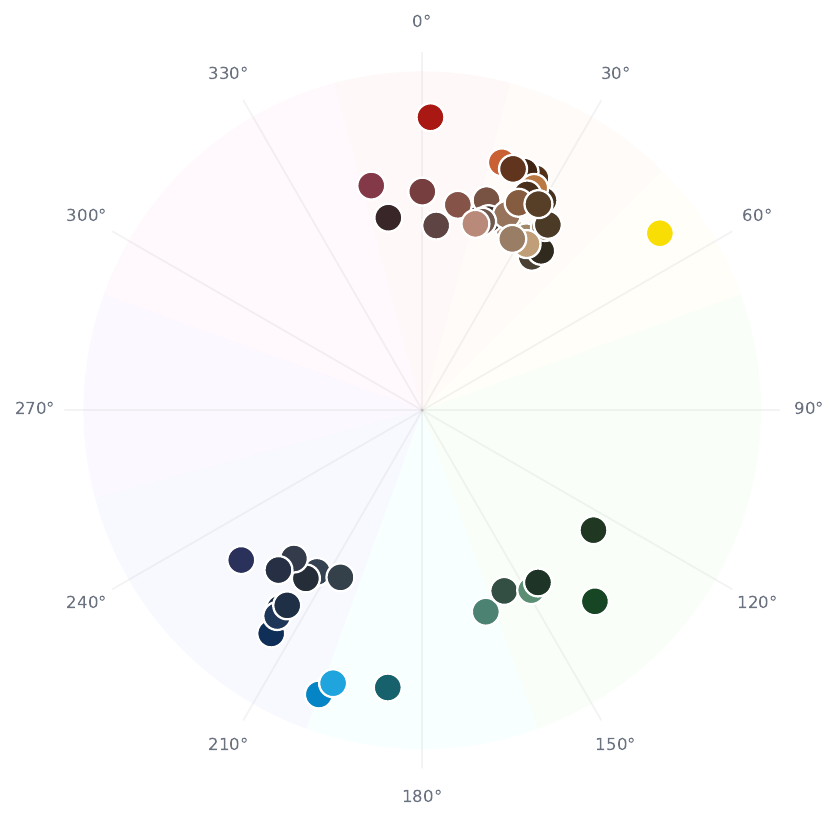}
  \caption{\rr{Accent-hue wheel: each of the 54 sites with a vivid brand accent placed at its accent hue, drawn in its own accent color. Accents concentrate in a warm/earth arc and a blue arc, with minor green and teal groups.}}
  \label{fig:palettewheel}
\end{figure}

\begin{figure}[h]
  \centering
  \includegraphics[width=\linewidth]{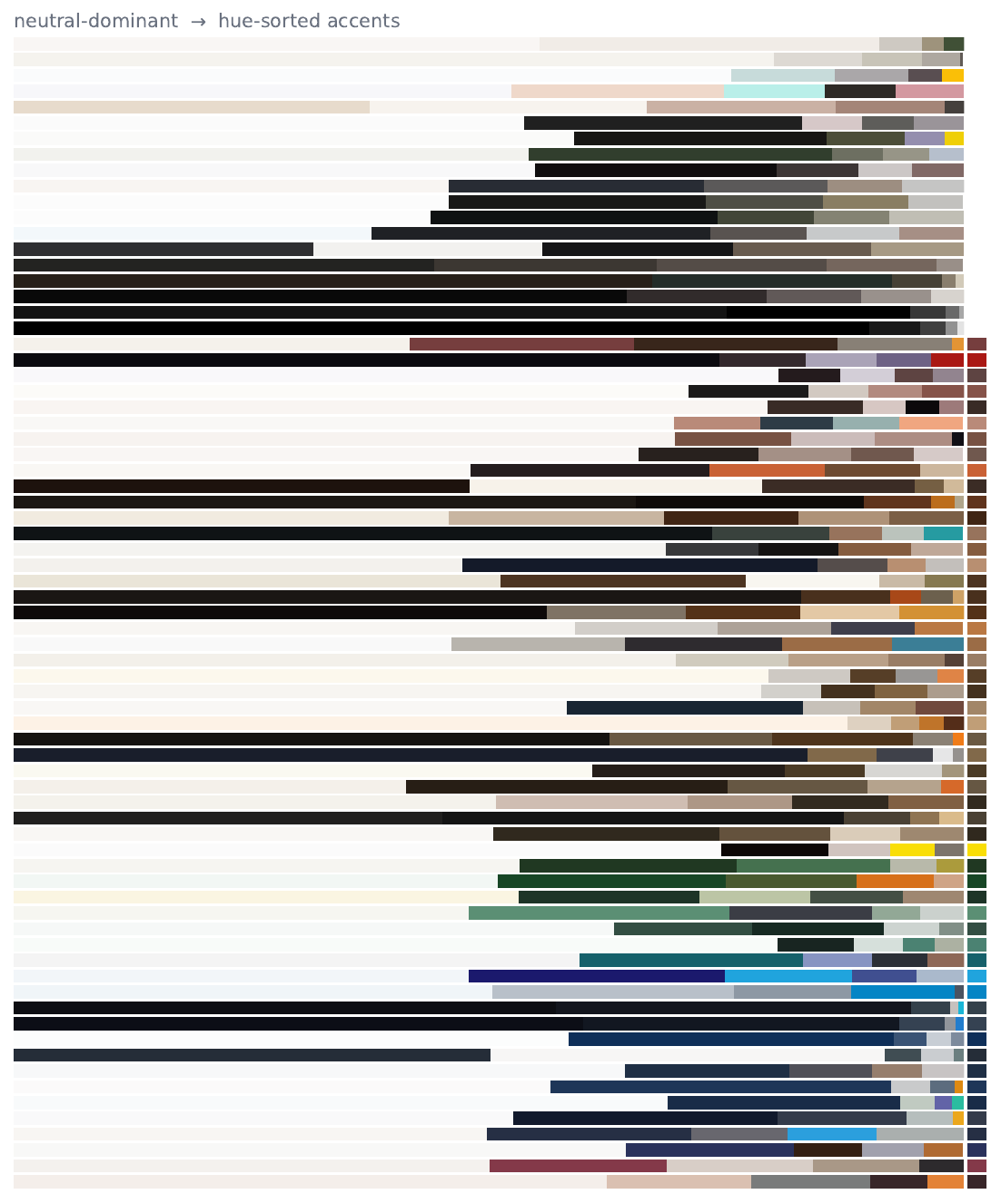}
  \caption{\rr{Palette wall: all 73 sites as area-weighted color strips (segment width $\propto$ screen area), ordered neutral-dominant first, then by accent hue; the right-edge tick is each site's brand accent. The large shared neutral base and the narrow warm/blue accent range are visible directly.}}
  \label{fig:palettewall}
\end{figure}

\rr{Palettes are overwhelmingly a neutral canvas plus a single accent: on average 63\% of screen area is neutral, and only 54 of 73 sites carry a vivid accent at all. Those accents collapse onto a narrow range (Figure~\ref{fig:palettewheel})---warm/earth tones (28) and trust-blue (11) dominate, with small green (5), teal (3), red (6), and yellow (1) minorities. The effective number of distinct palettes is strikingly small: the order-2 Vendi score on the full palette-similarity kernel is 2.2 (pulled down by the shared neutral base every site carries), and even restricting to the chromatic accent it is only $\approx$3 (inverse-Simpson 3.0 over hue families; Vendi 2.9 over a circular-hue kernel of the 54 accents). Color therefore converges even more sharply than layout ($\approx$12 distinct designs) or the input briefs (21.6 business domains). Figure~\ref{fig:palettewall} shows all 73 palettes as area-weighted strips, sorted from neutral-dominant to hue-sorted accents, making the shared neutral base and the warm/blue accent bands visible at a glance. Accent hue tracks genre only weakly (food and hospitality skew warm, technology skews blue) and cuts across the layout archetypes, consistent with the layout~$\times$~palette orthogonality of the preceding subsection.}

\subsection{\texorpdfstring{\rr{Typography and Its Convergence}}{Typography and Its Convergence}}

\begin{figure}[h]
  \centering
  \includegraphics[width=\linewidth]{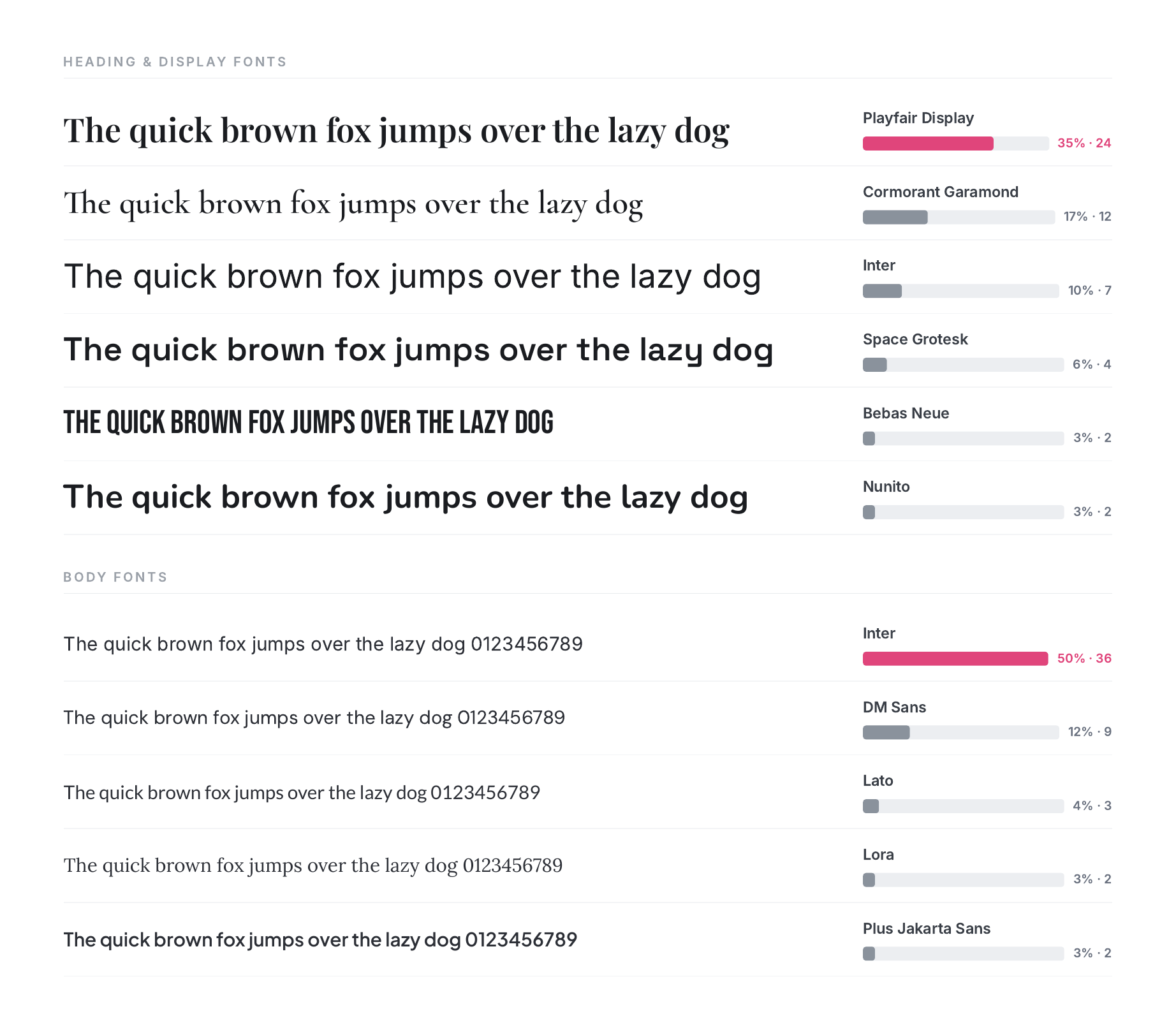}
  \caption{\rr{The cohort's type system. Dominant heading/display fonts (top) and body fonts (bottom) across the 73 sites, each shown set in its own typeface with the share of sites using it. A serif display heading (Playfair Display or Cormorant Garamond) over an Inter body recurs throughout; the same pangram is shown in every face.}}
  \label{fig:fonts}
\end{figure}

\rr{We extracted the rendered typography of every live site by loading it in headless Chromium and reading the computed \texttt{font-family} of its headings, body, and buttons (72 of the 73 sites loaded; one was unreachable at capture). Typography converges as sharply as color. Headings are dominated by two serif display faces---Playfair Display (24 sites, 35\%) and Cormorant Garamond (12, 17\%)---which together dress over half the cohort's headlines, while body text is dominated by a single sans, Inter (36, 50\%), with DM Sans second (9, 12\%). The three most common heading fonts cover 62\% of sites and the three most common body fonts 69\%, drawn from only about twenty distinct families each (40 across the whole cohort). The recurring pairing, a high-contrast serif display headline over an Inter or DM-Sans body, is the signature ``AI-generated'' type system; Figure~\ref{fig:fonts} renders the dominant faces set in themselves. Two caveats: a handful of sites specify only a generic stack (\texttt{system-ui}/\texttt{ui-sans-serif}) rather than a named webfont, and because all but one site loaded, the ``still-live'' subset is effectively the whole cohort.}

\subsection{\texorpdfstring{\rev{Cross-Model Probes}}{Cross-Model Probes}}

\rev{Two probes with non-DINO models contextualize what the measure captures; neither is a like-for-like robustness check, because no open non-DINO model currently provides a comparable style-focused global descriptor for images of this geometry. SigLIP~2 (an image-text encoder whose NaFlex variant preserves aspect ratio) agrees with the DINOv3 measure only moderately (per-site originality $r = \rr{.44}$), and its own similarity distribution saturates (mean pairwise similarity \rr{55.9\%}): an encoder trained to align images with text sees \rr{73} semantically near-identical small-business promotional pages, consistent with our measure capturing visual style beyond shared topic. Raw masked-autoencoder features (ViT-MAE), a reconstruction-trained model whose unfinetuned embeddings are documented to be weak global descriptors, show near-zero agreement (\rr{$r = .04$}) and are uninformative for this purpose. The decisive perceptual check remains a human pairwise-similarity validation on a stratified sample of pairs, planned for the archival version together with full documentation of the rendering procedure, since the elongated geometry of full-page screenshots sits outside the natural-image settings of the DINO evaluations \citep{oquab2023, simeoni2025}.}

\subsection{Environment}

Embeddings and clustering were produced with pinned dependency versions (umap-learn 0.5.11, hdbscan 0.8.41, transformers 5.1.0, torch 2.10.0), recorded in the archival package, because exact density-clustering partitions vary across library versions even at a fixed random seed.

\section{Stakeholder Feedback and Early Adoption}
\label{app:stakeholders}

The assignment required each student to present the site to at least one business stakeholder and one prospective customer, to document the feedback, and to iterate on it (Section~\ref{sec:task}). Sixty-five of the 78 written reflections describe this feedback in detail, and the accounts are informative in three ways: they show what real audiences asked of these sites, they record early adoption, and they show what nobody asked for.

\rev{First, the two audiences evaluated the sites through different lenses, and several students remark on the contrast explicitly. Owners treated the site as an operational tool and asked for accuracy and content: the owner of a coffee house asked for hours and location to be moved to the top of the page because ``they spend too much time on the phone answering these basic questions''; the stakeholder for a seafood market pushed on ``whether the wording matched how the market actually operates''; the owner of a cafe asked that a single ordering button be replaced with verified links to the delivery services customers actually use. Customers, in the same sessions, surfaced usability defects: a cart that could not be edited, a missing option to reschedule a booked session, a landing page a tester bluntly called unprofessional because text was layered over a background that already contained text. Feedback of both kinds was concrete, actionable, and taken: the reflections consistently describe returning to the tool and shipping the requested changes.}

\rev{Second, the deliverables were real enough to be adopted. The owner of a pizza restaurant ``shared it as the restaurant's official website.'' A wedding photographer was ``so impressed by the professional look and the ease of the admin portal that he is now planning to purchase the domain.'' The founder of a consultancy responded, ``I wish I could switch to this immediately,'' and asked for two additional service pages, which the student generated and shipped in the same session. Several other reflections describe owners planning to replace placeholder material with real content as a step toward going live. For businesses at the long tail of the web, these artifacts were not classroom exercises; some of them became, or are becoming, the business's public face.}

Third, and most relevant to this paper's argument, the recorded feedback polices content, trust, and function, and almost never position in the design distribution. Across the 65 accounts we find no request to look less like other websites. The clearest aesthetic redirection recorded is a photographer's stakeholder asking that the site feel ``more professional, modern, and inviting,'' which are the same generic gradient adjectives the process layer documents in prompts (Section~\ref{sec:thininput}). This is what the structural account predicts: owners and customers, like builders, see one artifact and not the distribution it belongs to, so market feedback corrects what a single viewer can verify and leaves genericity untouched. One reflection makes the loop fully circular: a student satisfied the feedback requirement by consulting ``structured AI personas'' representing an owner, executives, and three customer types, so that the artifact generated by the model was evaluated by simulations run on the same class of model.

\section{The Course Assignment}
\label{app:assignment}

For transparency, Appendix~\ref{app:assignment} reproduces a redacted course assignment and its full grading rubric. 

\subsection*{Overview}

In this assignment, you'll harness the power of AI-assisted development tools to create
a professional website for a real business. This exercise challenges you to act as both
designer and consultant, creating a functional web presence while gathering real-world
feedback from stakeholders and customers.

This is a practice for the future of work, where you'll be able to rapidly prototype
ideas, test them with real users, and iterate based on feedback, all within hours instead
of weeks.

\subsection*{Part 1: Business Selection}

Choose a real business that would benefit from a website.

\emph{Suggested options:}
\begin{itemize}
  \item A local business without a website (check Google Maps)
  \item A family business or business of friends/relatives
  \item Your own business or a business you're starting
  \item A small business with an outdated or poor website
  \item A Fortune 500 company's specific division/product that lacks strong web presence
  \item Your favorite coffee shop, restaurant, salon, boutique, charity, NGO, etc.
  \item A brand new tool/product/division that doesn't have a web presence yet
\end{itemize}

\emph{Requirements:}
\begin{itemize}
  \item Must be a legitimate, real business (not fictional)
  \item You must be able to contact someone associated with the business for feedback
  \item The business must have potential customers you can reach for feedback
\end{itemize}

\subsection*{Part 2: Website Creation}

\textbf{Tool recommendation:} Lovable (strongly recommended). You may also use any tool
of your choice (e.g., Claude Code, Codex, VSCode, Base44, etc.).

\emph{Website requirements} (examples of elements):
\begin{itemize}
  \item Homepage with a clear value proposition
  \item Business information: location, hours, contact details (whatever is relevant)
  \item Products/services section showcasing what the business offers
  \item Visual design that aligns with the business's brand and target audience
  \item Call-to-action (e.g., ``Contact Us,'' ``Visit Us,'' ``Order Now'')
  \item Mobile responsiveness (critical---must work beautifully on phones)
\end{itemize}

\emph{Technical requirements:}
\begin{itemize}
  \item Functional navigation between pages/sections
  \item Professional visual design with a consistent color scheme and typography
  \item Optimized images (fast loading)
  \item Clear information architecture
  \item Accessible and user-friendly interface
\end{itemize}

\emph{Quality standards:}
\begin{itemize}
  \item \textbf{Usability:} Can users easily find information and take desired actions?
  \item \textbf{Creativity:} Does it stand out while remaining professional?
  \item \textbf{Originality:} Does it avoid generic AI aesthetics?
  \item \textbf{Alignment:} Does it match the business's intended vibe and target audience?
\end{itemize}

\subsection*{Part 3: Testing \& Feedback (Critical)}

You must complete all of the following steps.

\textbf{3A: Business stakeholder feedback.} Show your website to at least one person who
works for or owns the business (owner, manager, employee, or family member). Ask
questions such as: What works well? What would you change? Does this accurately
represent the business? Would you actually use this website? Document their feedback and
make at least one iteration based on their input.

\textbf{3B: Potential customer feedback.} Show your website to at least one potential
customer of the business. Ask questions such as: Would this website make you want to
visit/use this business? Is it easy to find the information you'd need? Does it work well
on your phone? What would make you more likely to become a customer? Document their
feedback and compare it to the business stakeholder's feedback.

\textbf{3C: Cross-platform testing.} Test your website on at least two devices (a
desktop/laptop and a mobile phone) and two browsers, checking that all features work,
images load properly, text is readable, navigation functions correctly, and the site
looks professional on small screens.

\subsection*{Part 4: Recording Your Process (Optional Research Component)}

This is entirely optional. There is no penalty for opting out, and you do not need to
provide any justification. You are completely free to decide whether or not to
participate; no questions asked.

If you choose to participate, record the beginning of your website-building session
using Zoom (screen and audio, as you talk through your process).

\emph{Why we're doing this:} This helps us understand how people interact with AI tools
during creative work---how much time is spent evaluating AI-generated content, testing,
iterating, and correcting. This data contributes to research on human-AI collaboration
and ``vibe coding,'' a new and understudied area.

\emph{A note on confidentiality:} If your project involves sensitive or confidential
information (e.g., from your workplace), you have full flexibility. You can choose not to
record at all, or record only the portions that don't involve sensitive details.

\emph{Your recording will be:}
\begin{itemize}
  \item Anonymized and kept confidential
  \item Used only for research and educational purposes by the instructor and approved research team
  \item Not used for grading
  \item Stored securely according to IRB protocols
\end{itemize}

\emph{To opt out:} Simply don't submit a recording. That's it. Your grade will not be
affected in any way.

\subsection*{Part 5: Reflection Blog Post}

Write a 500+ word blog post reflecting on your experience creating a website for a real
business using AI tools. Your post should read like something you'd share on LinkedIn or
Medium: natural, engaging, and personal, while still clearly addressing the topics below.

\emph{What to cover (suggestions):}
\begin{itemize}
  \item \textbf{The business and your approach:} which business you chose, why, and your goals for the site.
  \item \textbf{Feedback and iteration:} who you got feedback from (at least one business stakeholder, one potential customer), what each liked and wanted changed, and what you changed in response.
  \item \textbf{Technical journey:} your three most important design decisions and why; an estimate, as a rough percentage, of how much of the final site came from AI-generated output versus your own direction; the biggest challenges you encountered and how you solved them.
  \item \textbf{Reflection and learning:} how well the site serves the business's goals, and what you learned about working with AI tools, web design, mobile-first design, and gathering stakeholder feedback.
\end{itemize}

\subsection*{Grading Rubric (15 Points Total)}

\begin{table}[h]
\centering
\begin{tabular}{p{4cm}p{1.2cm}p{7.5cm}}
\toprule
\textbf{Category} & \textbf{Points} & \textbf{Criteria} \\
\midrule
Website Quality \& Functionality & 4 pts & Fully functional website with all required elements; professional design; effective information architecture \\
\addlinespace
Mobile Responsiveness & 2 pts & Website works beautifully on mobile devices; properly tested on multiple devices \\
\addlinespace
Business Alignment \& Creativity & 2 pts & Website effectively represents the business's brand; creative and original design that avoids generic AI aesthetics \\
\addlinespace
Stakeholder Feedback \& Iteration & 2 pts & Gathered meaningful feedback from both business representative and potential customer; made thoughtful iterations based on feedback, well-documented in the blog post \\
\addlinespace
Reflection Quality & 5 pts & Thoughtful, detailed responses to reflection questions; clear insights about human-AI collaboration, design decisions, and learning outcomes; blog post reflects the student's voice \\
\bottomrule
\end{tabular}
\caption{Grading rubric for the assignment, reproduced in full.}
\label{tab:rubric}
\end{table}

\end{document}